\documentclass[aps,prresearch,twocolumn,superscriptaddress,longbibliography]{revtex4-2}
\usepackage[breaklinks=true,colorlinks,citecolor=blue,linkcolor=blue,urlcolor=blue]{hyperref}
\usepackage{amssymb,amsmath,amsfonts}
\usepackage{latexsym,multirow}
\usepackage{graphicx,dcolumn,bm}
\usepackage[usenames,dvipsnames]{xcolor}
\usepackage{array}
\usepackage{booktabs}

\begin{document}

\title{Chaos suppression via adaptive feedback control of intermittency:\\
From exactly solvable ergodic maps to interacting microbubble clusters
}

\author{Mohammad Yahyavi}
\email{mohammad.yahyavi@ntu.edu.sg}
\affiliation{Department of Physics, Bilkent University, TR-06800 Bilkent, Ankara, T\"{u}rkiye}

\author{Sina Gholizadeh}
\affiliation{Department of Physics, Bilkent University, TR-06800 Bilkent, Ankara, T\"{u}rkiye}

\author{Sohrab Behnia}
\email{S.behnia@sci.uut.ac.ir } 
\affiliation{Department of Physics, Urmia University of Technology, Urmia, Iran}

\author{Bilal Tanatar}
\email{tanatar@fen.bilkent.edu.tr}
\affiliation{Department of Physics, Bilkent University, TR-06800 Bilkent, Ankara, T\"{u}rkiye}

\date{\today}

\begin{abstract}
Intermittency represents a fundamental route to chaos in nonlinear dynamical systems. In this work we introduce an adaptive control strategy in which the control parameter of an intermittent system is promoted to a dynamical variable that evolves autonomously under an auxiliary nonlinear map drawn from the same functional hierarchy as the system itself. The construction eliminates the need for orbit identification, local linearization, and trajectory-triggered perturbations, which are central ingredients of conventional feedback schemes. The theoretical framework is developed within a class of one-dimensional nonlinear ergodic maps with exactly known invariant (Sinai--Ruelle--Bowen) measures, for which we derive in closed form (i) the dynamics and invariant measure of the evolving control parameter, (ii) the invariant measure of the coupled system, and (iii) the $q$-generalized Lyapunov exponents before and after control. The generalized Lyapunov spectrum serves as an analytical order parameter for the control process: the collapse of its positive regions provides a quantitative and initial-condition-independent signature of chaos suppression, and yields the sensitivity to initial conditions in explicit form. To establish the physical relevance of the approach beyond low-dimensional maps, we apply the same construction to a cluster of three interacting ultrasound-driven microbubbles described by the Keller--Herring model, promoting the experimentally accessible acoustic driving frequency to a dynamical variable. Systematic bifurcation and Lyapunov analyses, performed over wide ranges of driving pressure, frequency, and equilibrium radii, demonstrate that intermittent chaotic radial oscillations are progressively suppressed and replaced by stable periodic motion. These results establish intermittency regulation as a unifying and physically realizable mechanism for stabilizing complex nonlinear dynamics.
\end{abstract}

\maketitle

\section{Introduction}
\label{sec:intro}

Chaotic dynamics are characterized by the exponential divergence of initially nearby trajectories, quantified by a positive Lyapunov exponent $\lambda>0$ \cite{Dorfmanm,PRL1}. Among the various routes to chaos, intermittency occupies a central role, typically arising when a system parameter crosses a critical threshold \cite{IN-1m,IN-2m,IN-3m,NP1,rev1,rev2}. Intermittent dynamics are marked by long laminar phases interrupted by irregular chaotic bursts, with the specific form of intermittency determined by the underlying bifurcation mechanism. Prominent examples include the Pomeau--Manneville (PM) types \cite{IN-4m,IN-5m,Signal}, crisis-induced intermittency \cite{IN-6m}, and on--off intermittency \cite{IN-7m}.

Intermittent systems, both deterministic and noise-driven \cite{yhv2m,yhv3m}, exhibit universal scaling behavior that shares partial analogies with the Feigenbaum period-doubling scenario and PM intermittency \cite{IN-8m,IN-9m}. Such phenomena have been reported across a wide range of physical contexts, including the Lorenz system, periodically forced oscillators, Rayleigh--B\'enard convection, discrete nonlinear Schr\"odinger equations, and hydrodynamic turbulence \cite{INT-15-1m,INT-15-2m,INT-15-3m,INT-15-4m,MM-1m,NP2,CONT}. These observations underscore the fundamental and ubiquitous nature of intermittency in nonlinear dynamics.

Considerable efforts have therefore been devoted to controlling intermittent behavior. In PM intermittency, the stabilization of emerging periodic orbits has been explored as a control mechanism \cite{IN-15m}. For on--off intermittency, feedback strategies inspired by the Ott--Grebogi--Yorke (OGY) method have been proposed \cite{IN-11m,Sec4-10m}, while harmonic modulation has also been shown to reduce intermittent bursting \cite{IN-12m}. Crisis-induced intermittency has similarly been addressed through small parameter perturbations and boundary-crisis control techniques \cite{IN-14m,IN-10m}. Despite these advances, most existing approaches rely on precise state information, linearization near unstable periodic orbits, or carefully timed perturbations, which can limit their applicability in realistic systems.

Within the OGY framework \cite{IN-11m,Sec4-10m}, chaos is controlled by applying small perturbations to an accessible system parameter based on a local linearization of the Poincar\'e map near unstable periodic orbits. The method exploits the recurrence of chaotic trajectories, applying control only when the system returns to a neighborhood of a target orbit. While highly effective, it requires identification of unstable orbits, precise timing of perturbations, and real-time state monitoring, which can limit its applicability in complex or poorly characterized systems. Intermittency poses a particularly challenging problem for control, as rare but intense chaotic bursts dominate long-time dynamics and degrade predictability even when laminar phases are stable on average. Developing control strategies that suppress these bursts without disrupting the intrinsic nonlinear dynamics therefore remains an open problem.

In this work, we demonstrate that chaos can be effectively suppressed by directly regulating intermittency through an intrinsic dynamical mechanism. Specifically, we develop an adaptive feedback strategy in which the control parameter evolves autonomously according to an auxiliary nonlinear process, continuously modulating the system dynamics. This construction eliminates the need for orbit identification, local linearization, and trajectory-triggered perturbations, providing a fully autonomous and robust route to chaos suppression.

In conventional control approaches \cite{IN-11m,Sec4-10m,IN-12m,IN-14m,IN-10m,Sec4-11m,Sec4-12m,MM-2m}, the system is typically described by $\dot{x} = \mathcal{F}(x,\alpha)$, where $\alpha$ is treated as a static or externally tuned parameter. In contrast, in our approach, the dynamics of the controlled system are described by the state equation
\begin{equation}
	\dot{x} = \mathcal{F}(x,g(\alpha)),
\end{equation}
where $x \in \mathbb{R}^n$ denotes the system state and $g(\alpha) \in \mathbb{R}$ represents a changeable system parameter rather than a conventional external control input. Within the framework of ergodic theory of differentiable dynamical systems, discrete- and continuous-time descriptions are formally equivalent.

The control mechanism can be formulated conceptually as follows: (1) the system is described within a dynamical framework that captures its intrinsic nonlinear behavior; (2) the structures of the chaotic attractor, including unstable low-period orbits, are inherently encoded in the dynamics; (3) no explicit selection of a target orbit is required, as the control acts globally on the system; and (4) instead of applying localized perturbations when trajectories approach specific regions of phase space, the control parameter evolves autonomously in time and continuously modulates the system dynamics.

The purpose of the present, extended article is threefold. First, we give a complete and self-contained derivation of the adaptive control law within a hierarchy of one-parameter families of ergodic maps possessing exact invariant measures, and we display explicitly the dynamics of the evolving control parameter $\alpha_m$: its return map, fixed-point structure, chaotic trajectories, and exact invariant density (Sec.~\ref{sec:control}). Second, we clarify the role played by the $q$-generalized Lyapunov calculations. Conventional Lyapunov exponents average over the strongly heterogeneous laminar--burst structure of intermittent dynamics and are therefore insensitive to it; the $q$-generalized exponents, evaluated analytically with respect to the exact Sinai--Ruelle--Bowen (SRB) measure, act as the quantitative order parameter of the control process, and their collapse provides an initial-condition-independent, closed-form criterion for chaos suppression (Secs.~\ref{sec:maps} and \ref{sec:stat}). Third, we expand in full detail the application to a realistic physical system---a cluster of three interacting ultrasound-driven microbubbles described by the Keller--Herring model---presenting the complete set of bifurcation diagrams, Lyapunov spectra, time series, and phase-space portraits before and after control, together with all model parameters and numerical procedures required to reproduce the results (Sec.~\ref{sec:bubbles} and Appendix~\ref{app:numerics}).

The remainder of the paper is organized as follows. Section~\ref{sec:maps} introduces the hierarchy of ergodic maps, their exact invariant measures, the intermittency scenario of the representative map $\Phi_3(x)$, and the generalized Lyapunov exponents of the uncontrolled dynamics. Section~\ref{sec:control} develops the adaptive control construction, analyzes the explicit dynamics of the control parameter, and presents numerical results for the controlled map. Section~\ref{sec:stat} derives the invariant measure and the generalized Lyapunov exponent of the coupled (controlled) system and the resulting sensitivity to initial conditions. Section~\ref{sec:bubbles} applies the framework to interacting microbubble clusters. Section~\ref{sec:conclusions} summarizes our conclusions. Numerical methods are collected in Appendices~\ref{app:maps}-\ref{app:numerics}.

\section{Ergodic maps with exact invariant measures and intermittency}
\label{sec:maps}

\subsection{Hierarchy of one-parameter ergodic maps}
\label{sec:hierarchy}

Many canonical discrete-time models used in the study of chaos---including the logistic, H\'enon, tent, and Chirikov maps---can be unified within a hierarchy of one-parameter families of nonlinear maps with exact invariant measures \cite{jefm}. In particular, the logistic map is topologically conjugate to the first member of this hierarchy \cite{jefm}, providing a natural framework for analyzing chaotic dynamics while preserving their statistical properties.

The one-parameter families of chaotic maps of the interval $[0,1]$ with an exact invariant measure \cite{jefm} are constructed from Chebyshev polynomials of the first kind of degree $N$; their general closed form, together with the associated exact invariant (Sinai--Ruelle--Bowen) measures and the algebraic relations between the control parameter $\alpha$ and the measure parameter $\beta$, is collected in Appendix~\ref{app:maps}. The corresponding conjugate maps on $[0,+\infty)$ are given by \cite{jefm,jefm2,jefm3,jefm4,jefm5,jefm6}:
\begin{equation}\label{eq:til}
	\tilde{\Phi}_{N}(x)=h\circ \Phi_{N}(x)\circ h^{-1}
	=\frac{1}{\alpha ^{2}}\tan^{2}\!\left(N \arctan \sqrt{x}\right).
\end{equation}
Representative members of this family include
\begin{eqnarray}\label{eq:JEF}
	\begin{split}
		\Phi_{2}(x)&=\frac{\alpha^{2}(2x-1)^{2}}{4x(1-x)+\alpha^{2}(2x-1)^{2}},\\
		\Phi_{3}(x)&=\frac{\alpha^{2}x(4x-3)^{2}}{\alpha^{2}x(4x-3)^{2}+(1-x)(4x-1)^{2}},\\
		\Phi_{4}(x)&=\frac{\alpha^{2}\left[1-8x(1-x)\right]^{2}}{\alpha^{2}\left[1-8x(1-x)\right]^{2} +16 x(1-x)(1-2x)^{2}}.
	\end{split}
\end{eqnarray}
In particular, for $N=2$ the conjugate map takes the closed form
\begin{equation}\label{eq:S7}
	\tilde{\Phi}_{2}(x)=\frac{1}{\alpha^{2}}\tan^{2}\!\left(2\arctan\sqrt{x}\right)= \frac{1}{\alpha^2}\frac{4x}{(1-x)^2}.
\end{equation}
Each map of the hierarchy carries a positive parameter $\beta>0$, fixed by its exact invariant measure, which controls the statistical distribution of trajectories in phase space (Appendix~\ref{app:maps}). For the map of Eq.~\eqref{eq:S7} the two parameters are linked by $\alpha=2\beta/(1+\beta)$ [Eq.~\eqref{eq:S8}], so that $\tilde{\Phi}_{2}$ can be written entirely in terms of $\beta$ [Eq.~\eqref{eq:tildePhi2beta}]---a form that will play a central role in the control construction of Sec.~\ref{sec:control}.

\begin{figure}
	\includegraphics[width=\linewidth]{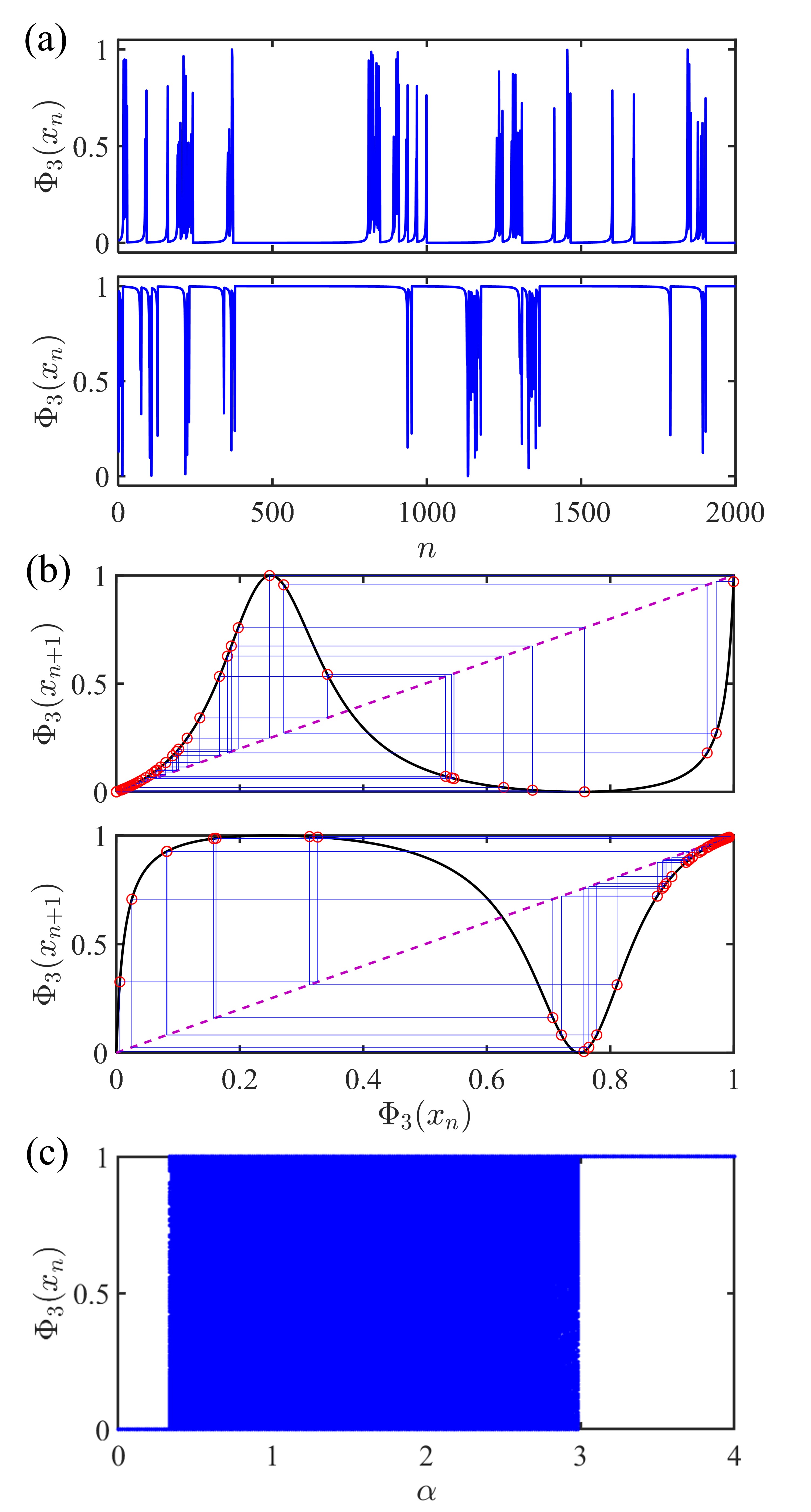}
	\caption{{\bf Intermittent dynamics in a one-dimensional nonlinear ergodic map.}
		(a)~Time series and (b)~cobweb diagrams of $\Phi_3(x_n)$ for representative values of the control parameter: (top) $\alpha=0.334$ and (bottom) $\alpha=2.999$, illustrating laminar phases interrupted by irregular bursts.
		(c)~Bifurcation diagram of $\Phi_3(x_n)$ as a function of $\alpha$. The blue region in the range $1/3<\alpha<3$ indicates the chaotic regime. For $\alpha<1/3$ ($\alpha>3$), the system converges to the stable fixed point $x=0$ ($x=1$).}
	\label{Fig.1}
\end{figure}

\subsection{Intermittency in the representative map $\Phi_3$}
\label{sec:intermittency}

We first illustrate the intermittency scenario within this hierarchy using the representative one-dimensional map $\Phi_{3}(x)$ of Eq.~\eqref{eq:JEF}, which exhibits intermittency without period-doubling or period-$n$-tupling cascades to chaos, and thus provides a minimal yet nontrivial framework for investigating laminar--burst dynamics and testing the proposed control mechanism.

Figure~\ref{Fig.1}(a) displays two typical time series of length $2000$ for $\alpha=0.334$ and $\alpha=2.999$, corresponding to two distinct regions within the chaotic regime. In both cases, the dynamics are characterized by long laminar phases interrupted by irregular chaotic bursts, reflecting the hallmark features of intermittency. For parameter values below the critical point, $\alpha<\alpha_c$, the system exhibits regular behavior associated with a stable fixed point in the Poincar\'e section. These properties make the map particularly suitable for isolating the role of intermittency in the transition to chaos and for assessing the efficiency of intermittency-based control strategies.

\begin{figure}
	\includegraphics[width=\linewidth]{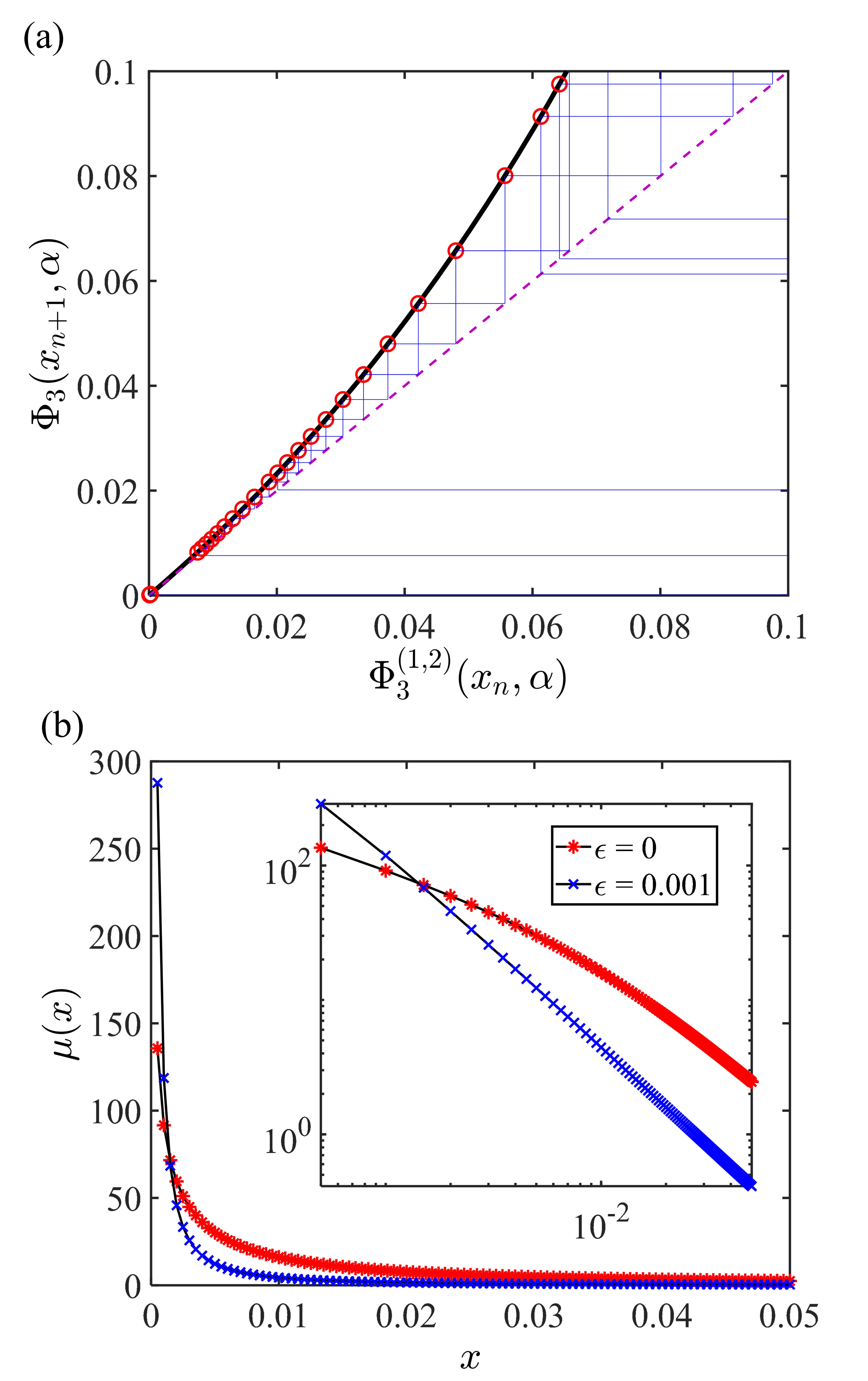}
	\caption{(a)~Enlargement of Fig.~\ref{Fig.1}(b) in the vicinity of $\Phi_3(x_{n})=0$. The cobweb diagram appears nearly tangent to the diagonal line, but it is not exactly tangent; trajectories therefore spend long laminar episodes inside the narrow channel before escaping in chaotic bursts.
	(b)~Invariant (SRB) measure of the coupled map $\Psi_3(x_{m},\alpha_{m})$ [Eq.~\eqref{eq:PS}] for $\epsilon=0$ and $\epsilon=0.001$, together with the same data on log-log scales. The symbols ($\ast$) and ($\times$) correspond, respectively, to the parameters used in Fig.~\ref{Fig.1}(a) and Fig.~\ref{Fig.2map}(a). The reshaping of the measure under control [discussed in Sec.~\ref{sec:invmeasure}] indicates the suppression of laminar dominance.}
	\label{Fig.SM}
\end{figure}

As $\alpha$ approaches the boundaries of the chaotic region ($\alpha-1/3=0.001$ and $3-\alpha=0.001$), the fixed point loses stability. The laminar dynamics are clearly revealed by the cobweb diagrams in Fig.~\ref{Fig.1}(b), where a narrow channel forms between the nonlinear map and the diagonal as $x^*\to0,1$. Trajectories evolving inside this channel give rise to laminar phases, while chaotic bursts occur upon escape, reflecting the presence of a ``ghost'' of the laminar region near an unstable periodic orbit. An enlargement of the near-tangency region is shown in Fig.~\ref{Fig.SM}(a): the graph of the map appears nearly tangent to the diagonal but is not exactly tangent, which is the geometric origin of the long laminar episodes. Consistently, the bifurcation diagram \cite{Bif} of $\Phi_3(x)$ shown in Fig.~\ref{Fig.1}(c) indicates that chaotic dynamics occur in the parameter range $1/3<\alpha<3$, where intermittent laminar--burst behavior emerges.

\subsection{Generalized Lyapunov exponents of the uncontrolled maps: motivation and derivation}
\label{sec:GLE-uncontrolled}

While the above dynamical picture clarifies the geometric origin of intermittency, a quantitative characterization requires tools sensitive to its strongly heterogeneous temporal structure. In intermittent systems, long laminar phases coexist with rare chaotic bursts, rendering conventional Lyapunov exponents inadequate, as they capture only the average exponential divergence of trajectories. This point deserves emphasis, since it defines the role played by the Lyapunov calculations throughout this paper: the conventional exponent $\lambda_1$ is a single time-averaged number that cannot distinguish a trajectory spending most of its time in a laminar channel (with rare violent bursts) from one that is uniformly weakly chaotic; near the intermittency threshold it vanishes with a nonanalytic scaling and loses discriminating power precisely where the control problem is most interesting. To overcome this limitation, we characterize the dynamics using the $q$-generalized Lyapunov exponent \cite{Sec3_3-2m,Sec3_3-3m,Sec3_3-4m,Sec3_3-5m,Sec3_3-1m,Tsallis98}, which resolves the fluctuations of finite-time stretching rates and thereby provides a natural diagnostic for the scaling properties of intermittent chaos. Because the maps of Sec.~\ref{sec:hierarchy} possess exact SRB measures, all $q$-generalized exponents can be evaluated {\it analytically} and independently of initial conditions; in Sec.~\ref{sec:stat} the same quantity, recomputed for the controlled system, will serve as the order parameter that quantifies the efficiency of the control.

Unlike the standard Lyapunov exponent---which is defined using the natural logarithm and quantifies the average exponential divergence of nearby trajectories---the generalized Lyapunov exponent replaces the ordinary logarithm by the $q$-logarithm, $\log_q x=(x^{1-q}-1)/(1-q)$. This leads to a one-parameter family of instability measures, which reduces to the conventional Lyapunov exponent in the limit $q \to 1$, while providing a more general characterization of dynamical sensitivity for arbitrary values of $q$. For a map admitting an invariant density $\mu(x)$, the generalized Lyapunov exponent is defined as \cite{Sec3_3-2m,Sec3_3-3m,Sec3_3-4m,Sec3_3-5m,Sec3_3-1m}
\begin{equation}\label{eq:GLSM}
	\lambda_{q}\big(\mu,\Phi_N(x)\big)
	=
	\int \mu(x)\,dx\;
	\log_{q}\left|
	\frac{d}{dx}\Phi_N(x)
	\right|,
\end{equation}
with $\mu(x)$ the invariant SRB measure of Eq.~\eqref{eq:Mes}. The generalized Lyapunov exponent is a topological invariant preserved under the conjugacy map,
\begin{equation}
	\lambda_{q}\big(\mu,\Phi_N(x)\big)=\lambda_{q}\big(\tilde{\mu},\tilde{\Phi}_N(x)\big).
\end{equation}
Because the SRB measures of the maps $\Phi_N$ are known exactly [Eq.~\eqref{eq:Mes}], the integral in Eq.~\eqref{eq:GLSM} can be carried out in closed form for arbitrary $N$. The complete derivation, based on the polynomial representation of the Chebyshev maps and the algebraic relation between $\alpha$ and $\beta$, is presented in Appendix~\ref{app:GLEun}; the general result is quoted there in Eq.~\eqref{eq:SMGL}.
For the representative case $\Phi_3(x)$, the general result reduces to \cite{jefm,Behniam}
\begin{equation}
	\lambda_q(\alpha)
	=\log_q\!\left[
	\frac{3(1+\beta+2\sqrt{\beta})^2}
	{(1+3\beta)(3+\beta)}
	\right],\label{eq:GLE}
\end{equation}
with $\beta>0$, where the parameter $\beta$ is related to the control parameter $\alpha$ through the algebraic relation $\alpha = (3\beta+1)/(\beta+3)$ [Eq.~\eqref{eq:alphabeta}], which enables $\lambda_q$ to be expressed equivalently as a function of either $\beta$ or $\alpha$.

The behavior of the generalized Lyapunov exponent near the transition to chaos via intermittency, corresponding to the limit $\beta \to 0$, is described by
\begin{equation}
	\left\{
	\begin{array}{l}
		\lambda_{q}\big(\mu,\Phi_{N}(x,\alpha=N+0^{-})\big)\sim
		\dfrac{(N-\alpha)^{1/2}}{\ln q},\\[10pt]
		\lambda_{q}\big(\mu,\Phi_{N}(x,\alpha=\tfrac{1}{N}+0^{+})\big)\sim
		\dfrac{(\alpha-\tfrac{1}{N})^{1/2}}{\ln q}.
	\end{array}\right.
	\label{eq:scaling}
\end{equation}
Pomeau and Manneville \cite{IN-4m,IN-5m} assumed a uniform reinjection probability into the laminar phase and calculated the scaling behavior of both the average laminar length and the Lyapunov exponent for the three types of intermittent chaos; the scaling of Eq.~\eqref{eq:scaling} provides an analogous signature of intermittency for the present family of maps. For inverse control parameters $q>1$, the generalized Lyapunov exponent vanishes at the transition and exhibits a linear dependence on $q$, reflecting the dominance of laminar phases in the dynamics. Although intermittency has been widely studied, its characterization through generalized Lyapunov exponents remains comparatively unexplored.

\section{Adaptive feedback control via a dynamical control parameter}
\label{sec:control}

\subsection{General construction and control law}
\label{sec:construction}

We construct an adaptive control framework for nonlinear dynamical systems exhibiting intermittency by embedding the control mechanism directly into the system dynamics. The controlled system is defined by
\begin{equation}
	x_{m+1}=\Phi_N\!\left(x_m,g(\alpha_m)\right),
	\label{eq:controlled}
\end{equation}
where $\Phi_N$ is a nonlinear map of the hierarchy of Sec.~\ref{sec:hierarchy} and $\alpha_m$ is the control parameter promoted to a dynamical variable. The control law in Eq.~\eqref{eq:controlled} is not an ad hoc modification, but follows from a dynamical consistency principle. The central idea is to let $\alpha_m$ evolve according to an auxiliary reference map,
\begin{equation}
	\alpha_{m+1}=R(\alpha_m),
	\label{eq:refmap}
\end{equation}
whose dynamics are drawn from the same functional hierarchy as the system itself, and to embed the evolution of $\alpha_m$ consistently into the system dynamics.

To make this construction explicit, the reference map is selected in accordance with the structure of the conjugate dynamics, Eq.~\eqref{eq:tildePhi2beta}, yielding
\begin{equation}
	R(\alpha)=\left(\frac{1+\beta}{\beta}\right)^{\!2}\frac{\alpha}{(1-\alpha)^2}.
	\label{eq:R}
\end{equation}
We aim to replace the static control parameter $\alpha$ with a dynamical one that follows the reference map $R$. To this end, we introduce a reparametrization $\eta(\alpha)$ and require that the {\it effective} parameter in the conjugate map follows the same reference dynamics. This condition can be expressed as
\begin{equation}
	\frac{1}{g(\alpha_m)^2}
	\equiv
	\frac{\eta(\alpha_{m+1})}{\eta(\alpha_m)}
	\frac{1}{R(\alpha_m)^2}.
	\label{eq:consistency}
\end{equation}
The control law $\alpha_{m+1}=R(\alpha_m)$ is autonomous and constitutes the core of the construction. In this formulation, the step $(m+1)$ is directly related to the current step $m$ through the reference dynamics $R$, while being rescaled by the factor $\eta(\alpha_{m+1})/\eta(\alpha_m)$, which provides a mechanism to tune the coupling strength. The system inherently ``adapts,'' since the trajectory of the control parameter is predetermined to evolve within the same functional class as the original map; the coupling strength controls how strongly this predefined trajectory influences the main system through the function $g(\alpha)$.

Solving Eq.~\eqref{eq:consistency}, the effective control function for the target system is obtained as
\begin{equation}
	\label{eq:g}
	g(\alpha_m)=\frac{2\eta(\alpha_m)}{1+\eta(\alpha_m)}
	\sqrt{\frac{\eta(\alpha_{m+1})}{\eta(\alpha_m)}},
\end{equation}
where the evolution of the control parameter is given by Eqs.~\eqref{eq:refmap} and \eqref{eq:R}, which establishes an exact dynamical synchronization between the auxiliary map and the controlled system. Finally, we take
\begin{equation}
	\eta(\alpha)=(1+\epsilon \alpha)^2,
	\label{eq:eta}
\end{equation}
where $\epsilon \in [0,1]$ controls the strength of the dynamical coupling and tunes the extent of the chaotic region.

This construction demonstrates that $g(\alpha)$ is not an arbitrary choice, but rather acts as a dynamical weight determined by the underlying mapping. The function $\eta(\alpha)$ serves as an amplitude modulator tuned by $\epsilon$, while the ratio $\sqrt{\eta(\alpha_{m+1})/\eta(\alpha_m)}$ encodes the forward-time evolution from step $m$ to $m+1$ and injects it into the system dynamics; the prefactor $2\eta/(1+\eta)$ ensures that this response remains bounded. As a result, $g(\alpha)$ enforces continuous adaptation of the system to the evolving control-parameter trajectory. The resulting coupled dynamics can be written compactly as
\begin{equation}
	\label{eq:PsiN}
	\Psi_{N}(x_m,\alpha_m)=
	\left\{
	\begin{aligned}
		x_{m+1} &= \Phi_N\!\left(x_m,g(\alpha_m)\right),\\
		\alpha_{m+1} &=
		\left(\frac{1+\beta}{\beta}\right)^{\!2}
		\frac{\alpha_m}{(1-\alpha_m)^2},
	\end{aligned}
	\right.
\end{equation}
where $x_m$ and $\alpha_m$ denote the state variable and the dynamical control parameter at iteration $m$, respectively. This construction replaces externally applied perturbations with an intrinsic dynamical mechanism, forming a closed feedback loop between the system state and the control parameter.

\subsection{Explicit dynamics of the control parameter}
\label{sec:alphadyn}

Because the entire control mechanism rests on the autonomous evolution of $\alpha_m$, we now analyze the reference map $R(\alpha)$ of Eq.~\eqref{eq:R} explicitly. The map $R$ coincides, by construction, with the conjugate map $\tilde\Phi_2$ of Eq.~\eqref{eq:tildePhi2beta}: it is therefore a {\it fully chaotic ergodic map on the half line} $[0,\infty)$, smoothly conjugate to $\Phi_2(x)$ on $[0,1]$ [and, for $\beta=1$, to the logistic map at its fully chaotic point]. Its relevant properties are as follows.

\begin{figure*}
	\includegraphics[width=\textwidth]{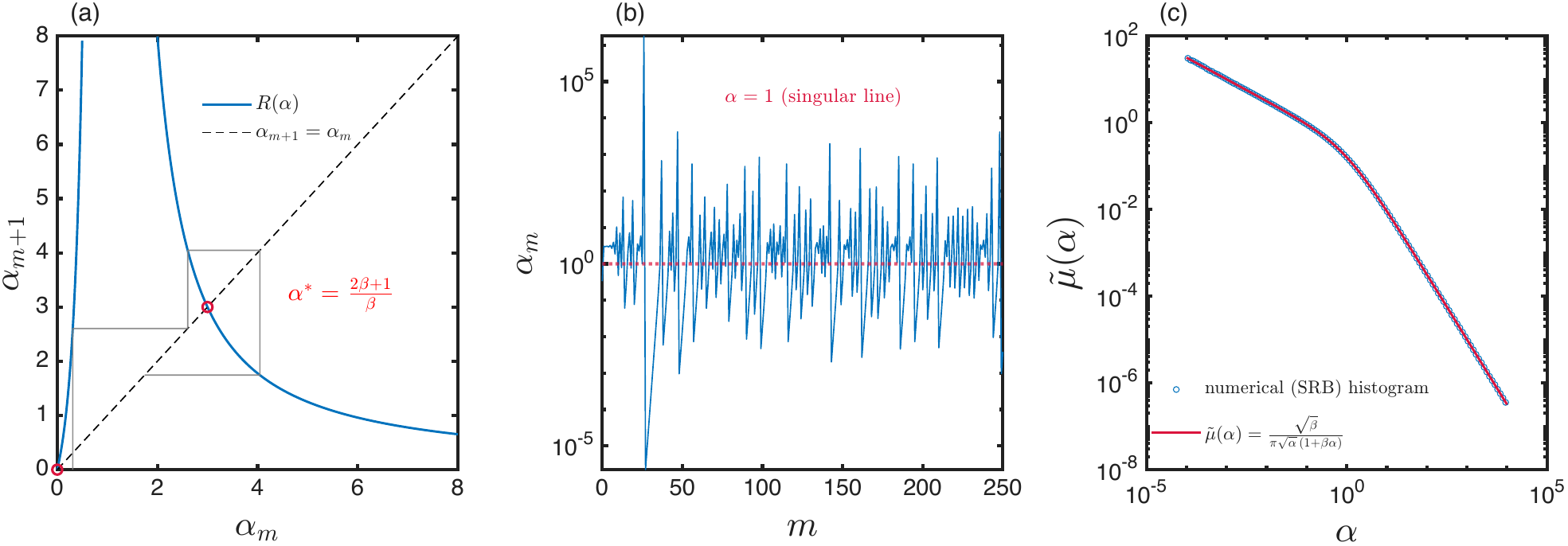}
	\caption{{\bf Explicit dynamics of the control parameter $\alpha_m$} under the reference map $R(\alpha)=[(1+\beta)/\beta]^{2}\,\alpha/(1-\alpha)^{2}$ [Eq.~\eqref{eq:R}], shown for $\beta=1$, i.e., $R(\alpha)=4\alpha/(1-\alpha)^{2}$.
	(a)~Return map (blue) with the identity line (dashed) and a cobweb trajectory (gray) started at $\alpha_0=0.31$; the nontrivial fixed point $\alpha^{*}=(2\beta+1)/\beta=3$ [Eq.~\eqref{eq:alphafixed}] is unstable.
	(b)~Time series $\alpha_m$ (logarithmic scale) for $\alpha_0=0.334$; the trajectory is chaotic, repeatedly approaching the singular line $\alpha=1$ (dotted) and being reinjected to large values.
	(c)~Histogram of $10^{7}$ iterates (circles) compared with the exact invariant density $\tilde\mu(\alpha)=\sqrt{\beta}/[\pi\sqrt{\alpha}(1+\beta\alpha)]$ [Eq.~\eqref{eq:mualpha}] (solid line) on log-log scales. All panels are generated directly from Eq.~\eqref{eq:R} with double-precision iteration and a transient of $10^{3}$ steps.}
	\label{Fig.alpha}
\end{figure*}

{\it Fixed points and instability.---}The fixed points of Eq.~\eqref{eq:R} satisfy $(1-\alpha^{*})^{2}=\left[(1+\beta)/\beta\right]^{2}$, giving
\begin{equation}
	\alpha^{*}_{0}=0,\qquad
	\alpha^{*}=\frac{2\beta+1}{\beta},
	\label{eq:alphafixed}
\end{equation}
(the third root, $\alpha^{*}=-1/\beta$, lies outside the physical domain). Both fixed points are unstable: $R'(0)=[(1+\beta)/\beta]^{2}>1$, and a direct computation gives $|R'(\alpha^{*})|>1$ for all $\beta>0$. Consequently, the control-parameter trajectory never settles on a fixed value; instead it explores the half line ergodically, repeatedly approaching the singular line $\alpha=1$ (where the map diverges) and being reinjected to large values, as shown in Figs.~\ref{Fig.alpha}(a) and \ref{Fig.alpha}(b).

{\it Exact invariant density.---}Since $R=\tilde\Phi_2$, the invariant density of the $\alpha$ dynamics follows from Eq.~\eqref{eq:Mes} by the conjugacy transformation $h(x)=(1-x)/x$ and reads
\begin{equation}
	\tilde{\mu}(\alpha)=\frac{\sqrt{\beta}}{\pi\sqrt{\alpha}\,(1+\beta\alpha)},
	\qquad \alpha\in[0,\infty).
	\label{eq:mualpha}
\end{equation}
Figure~\ref{Fig.alpha}(c) compares Eq.~\eqref{eq:mualpha} with a histogram of $10^{7}$ iterates of Eq.~\eqref{eq:R}: the agreement confirms both the ergodicity of the $\alpha$ dynamics and the analytical form of its SRB measure. The density diverges as $\alpha^{-1/2}$ at small $\alpha$ and decays as $\alpha^{-3/2}$ at large $\alpha$; the control parameter therefore spends most of its time at small values, punctuated by excursions to large $\alpha$, mirroring the laminar--burst structure of the system it controls.

{\it Consequence for the coupled system.---}It is precisely this statistically stationary, ergodic evolution of $\alpha_m$---rather than convergence of $\alpha_m$ to a fixed value---that underlies the control mechanism. The effective parameter entering the system is $g(\alpha_m)$ of Eq.~\eqref{eq:g}: through the modulator $\eta(\alpha)=(1+\epsilon\alpha)^2$, the excursions of $\alpha_m$ are converted into a bounded, continuously varying modulation of the map nonlinearity, whose statistics are fixed by Eq.~\eqref{eq:mualpha}. Increasing $\epsilon$ increases the weight of these modulations and progressively deforms the invariant measure of the coupled system [Sec.~\ref{sec:invmeasure} and Fig.~\ref{Fig.SM}(b)], confining trajectories near the laminar channel and suppressing bursts. We emphasize that the bifurcation diagrams of the controlled system discussed below are plotted as functions of the dynamical parameter $\alpha_m$, which evolves according to Eq.~\eqref{eq:R} and is independent of the coupling strength $\epsilon$; the parameter $\epsilon$ solely controls the strength of the dynamical coupling, while $\alpha_m$ governs the intrinsic map dynamics.

\subsection{Controlled map $\Psi_3$: numerical results}
\label{sec:controlledmap}

Within this framework, the control scheme can be explicitly realized for the representative case $N=3$, yielding the coupled map
\begin{equation}
	\label{eq:PS}
	\Psi_{3}(x_m,\alpha_m)=
	\begin{cases}
		x_{m+1} =
		\dfrac{\mathcal{N}(x_m,\alpha_m)}
		{\mathcal{N}(x_m,\alpha_m)+(1-x_m)(4x_m-1)^2},\\[8pt]
		\alpha_{m+1} =
		\left(\dfrac{1+\beta}{\beta}\right)^{\!2}
		\dfrac{\alpha_m}{(1-\alpha_m)^2},
	\end{cases}
\end{equation}
where $\mathcal{N}(x_m,\alpha_m)=g(\alpha_m)^2\, x_m(4x_m-3)^2$. In this formulation, feedback control emerges intrinsically through the dynamical evolution of the control parameter itself. The auxiliary chaotic map governing $\alpha_m$ continuously modulates the system dynamics, confining trajectories near the invariant subspace and suppressing intermittent bursts. Importantly, this mechanism operates without external perturbations or explicit state monitoring, providing a robust and adaptive route for controlling intermittency in low-dimensional chaotic systems. The dependence on the coupling parameter $\epsilon$ explicitly quantifies the influence of the auxiliary dynamics on the stability of the controlled system.

\begin{figure}
	\includegraphics[width=\linewidth]{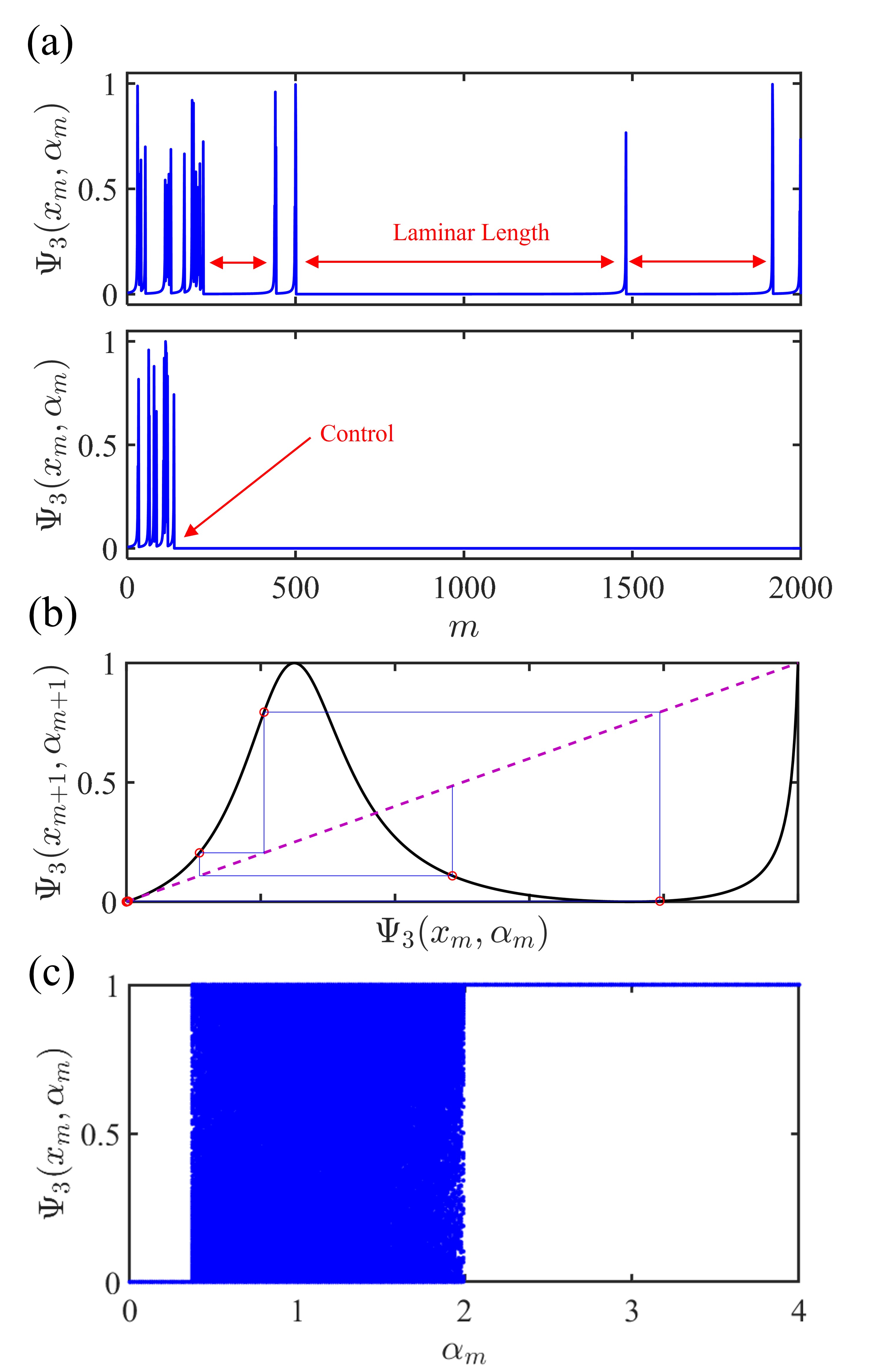}
	\caption{{\bf Intermittency control in one-dimensional nonlinear ergodic maps.}
		(a)~Time series of the coupled map $\Psi_3(x_m,\alpha_m)$ for $\alpha=0.334$, shown for two different coupling strengths:
		(top) $\epsilon=0.001$, exhibiting pronounced intermittent bursts; and
		(bottom) $\epsilon=0.003$, where laminar phases are significantly enhanced.
		(b)~Corresponding cobweb diagram of the intermittent dynamics for the same parameters as in panel (a) (top), illustrating the laminar channel structure responsible for intermittency.
		(c)~Bifurcation diagram of $\Psi_3(x_m,\alpha_m)$ for $\epsilon=0.1$, demonstrating that increasing the coupling strength progressively suppresses chaotic dynamics and stabilizes the system. The abscissa is the dynamical parameter $\alpha_m$, which evolves according to Eq.~\eqref{eq:R} and is independent of $\epsilon$.}
	\label{Fig.2map}
\end{figure}

It should be emphasized that the appropriate choice of the coupling strength $\epsilon$ depends on the specific nature of the system under study. As illustrated in Figs.~\ref{Fig.2map}(a) and \ref{Fig.2map}(b), increasing $\epsilon$ leads to a progressive enhancement of laminar phases and a corresponding reduction of intermittent bursts. For sufficiently large values of $\epsilon$, chaotic dynamics are fully suppressed, yielding stable behavior, as shown in Fig.~\ref{Fig.2map}(c). This result demonstrates that regulating intermittent dynamics provides a direct and systematic route for suppressing chaotic behavior, thereby establishing intermittency control as an effective mechanism for chaos control.

\section{Statistical characterization of the controlled dynamics}
\label{sec:stat}

\subsection{Invariant measure of chaotic maps with a dynamical parameter}
\label{sec:invmeasure}

The invariant measure characterizes the statistical properties of a dynamical system and is closely related to fundamental features such as ergodicity, the generalized Lyapunov exponent, and entropy. Mathematical analyses \cite{24,25,26,27,m1,m2} have shown that physically relevant invariant measures may exist in different forms: one concentrated on marginally stable fixed points, and another distributed continuously over the interval $[0,1]$. The former still possesses well-defined properties of SRB measures and yields a probability density function in the form of a distribution localized at the marginally stable fixed points.

If the invariant measure $\mu(x)$ does not depend on the initial condition $x_{0}$, the system is said to be ergodic. In general, however, the exact determination of invariant measures for dynamical systems is a highly nontrivial problem. Analytical expressions are known only for a limited class of systems, such as one-parameter families of one-dimensional piecewise linear maps \cite{28,29,30}, including the Baker and tent maps, or for unimodal maps such as the logistic map at specific values of the control parameter.

For a deterministic map such as $\Phi_{N}(x)$, invariance of the measure $\mu(x)$ is expressed by the Frobenius--Perron (FP) integral equation, whose solution is the eigenfunction of the FP operator with eigenvalue unity \cite{Dorfmanm,32}. For the coupled map $\Psi_N(x_m,\alpha_m)$ of Eq.~\eqref{eq:PsiN}, the invariant measure satisfies a two-dimensional generalization of the FP equation. Because the evolution of $\alpha_m$ is autonomous---the Jacobian factor $\partial\alpha_m/\partial\alpha_{m+1}$ does not depend on $x_m$---the invariant measure factorizes as
\begin{equation}
	\mu(x,\alpha)\,dx\,d\alpha=\big[\mu(x \mid \alpha)\,dx\big]\big[\mu(\alpha)\,d\alpha\big],\label{eq:B7}
\end{equation}
where $\mu(\alpha)$ denotes the invariant measure of the dynamical parameter $\alpha$, and $\mu(x \mid \alpha)$ represents the conditional measure of the dynamical system for a fixed value of $\alpha$. This decomposition reflects the fact that $\alpha$ is not deterministic in the ordinary sense, but evolves according to its own autonomous dynamics.

Solving the two-dimensional FP equation with a conditional-measure ansatz---the complete derivation, including the preimage structure of the coupled map, the associated Jacobians, and the verification of the ansatz, is presented in Appendix~\ref{app:measure}---we obtain the conditional measure
\begin{equation}
	\mu(x\mid\alpha)=\frac{1}{\pi}\frac{\sqrt{\eta(\alpha)}}{\sqrt{x(1-x)}\,\big[\eta(\alpha)+(1-\eta(\alpha))x\big]},\label{eq:B16}
\end{equation}
with $\eta(\alpha)$ given by Eq.~\eqref{eq:eta}. The parameter measure $\mu(\alpha)$ requires no separate display: it is precisely the invariant density of the reference map already obtained in Sec.~\ref{sec:alphadyn}, i.e., the conjugate transform of Eq.~\eqref{eq:mualpha} back to the unit interval, and has the same functional form as the SRB measure of the uncontrolled hierarchy [Eq.~\eqref{eq:Mes}] with $x\to\alpha$; its explicit expression is given in Eq.~\eqref{eq:B17} of Appendix~\ref{app:measure}.

Figure~\ref{Fig.SM}(b) shows the resulting invariant measure of the coupled system. In the uncontrolled case ($\epsilon=0$), the measure accumulates strongly near the laminar region, reflecting intermittent trapping. When control is applied ($\epsilon=0.001$), the invariant measure is significantly reshaped, indicating suppression of laminar dominance and reduction of chaotic bursts. These results demonstrate that intermittency control acts directly on the underlying phase-space structure of the dynamics.

\subsection{Generalized Lyapunov exponent of the controlled system}
\label{sec:GLE-controlled}

To quantitatively assess the efficiency of the control process, we again employ the generalized Lyapunov exponent, now evaluated with respect to the invariant measure of the coupled system derived in Sec.~\ref{sec:invmeasure}. This is the precise sense in which the Lyapunov calculations enter the control problem: the same analytically accessible quantity that diagnoses intermittency in the uncontrolled map [Eq.~\eqref{eq:GLE}] becomes, after control, an explicit function of the coupling strength $\epsilon$, so that the closure or persistence of positive-$\lambda_q$ regions can be read off in closed form and compared before and after control without any dependence on initial conditions or trajectory sampling.

Using Eqs.~\eqref{eq:B16} and \eqref{eq:B17}, the generalized Lyapunov exponent can be written as the ensemble average
\begin{equation}
	\lambda_{q}(\mu,\Psi_{N})=\int d\alpha\, \mu(\alpha)\,
	\lambda_{q}(\mu,\Psi_{N}\mid \alpha),
\end{equation}
where $\lambda_{q}(\mu,\Psi_{N}\mid \alpha)$ denotes the generalized Lyapunov exponent for a fixed value of the parameter $\alpha$. This expression is equivalent to
\begin{equation}\label{eq:lambda1}
	\lambda_{q}(\mu,\Psi_{N}) =\int \mu(\alpha)\, d\alpha\int dx\,\mu(x\mid
	\alpha)\,\log_{q}\left|\frac{d}{dx}\Psi_{N}(x)\right|.
\end{equation}
Carrying out the double integral in the conjugate representation of Eq.~\eqref{eq:til}---using the preimage sums and trigonometric integral identities of Refs.~\cite{jefm,jefm2,jefm3,jefm4,jefm5,jefm6}; the full derivation is given in Appendix~\ref{app:GLEc}---the generalized Lyapunov exponent of $\Psi_{3}(x_m,\alpha_m)$ is obtained in closed form as
\begin{equation}\label{eq:GLE2}
	\lambda_{q}(\epsilon)=\log_{q}\big[\Gamma(\beta,\epsilon)\big],
\end{equation}
where
\begin{equation}\label{eq:GLE32}
	\Gamma(\beta,\epsilon)=
	\frac{
		3\left(\sqrt{2}+\sqrt{\epsilon/\beta}\right)^{8}
	}{
		\left(
		\frac{2\beta+\epsilon}{\beta}
		+
		\sqrt{\frac{6\epsilon}{\beta}}
		\right)^{2}
		\left(
		\frac{2\beta+\sqrt{3}\,\epsilon}{\beta}
		+
		\sqrt{\frac{(6+4\sqrt{3})\epsilon}{\beta}}
		\right)^{2}
	}.
\end{equation}
Considering Eqs.~\eqref{eq:GLE2} and \eqref{eq:GLE32}, it is evident that intermittency can be effectively controlled by choosing appropriate values of the coupling strength $\epsilon$.

\begin{figure}
	\includegraphics[width=\linewidth]{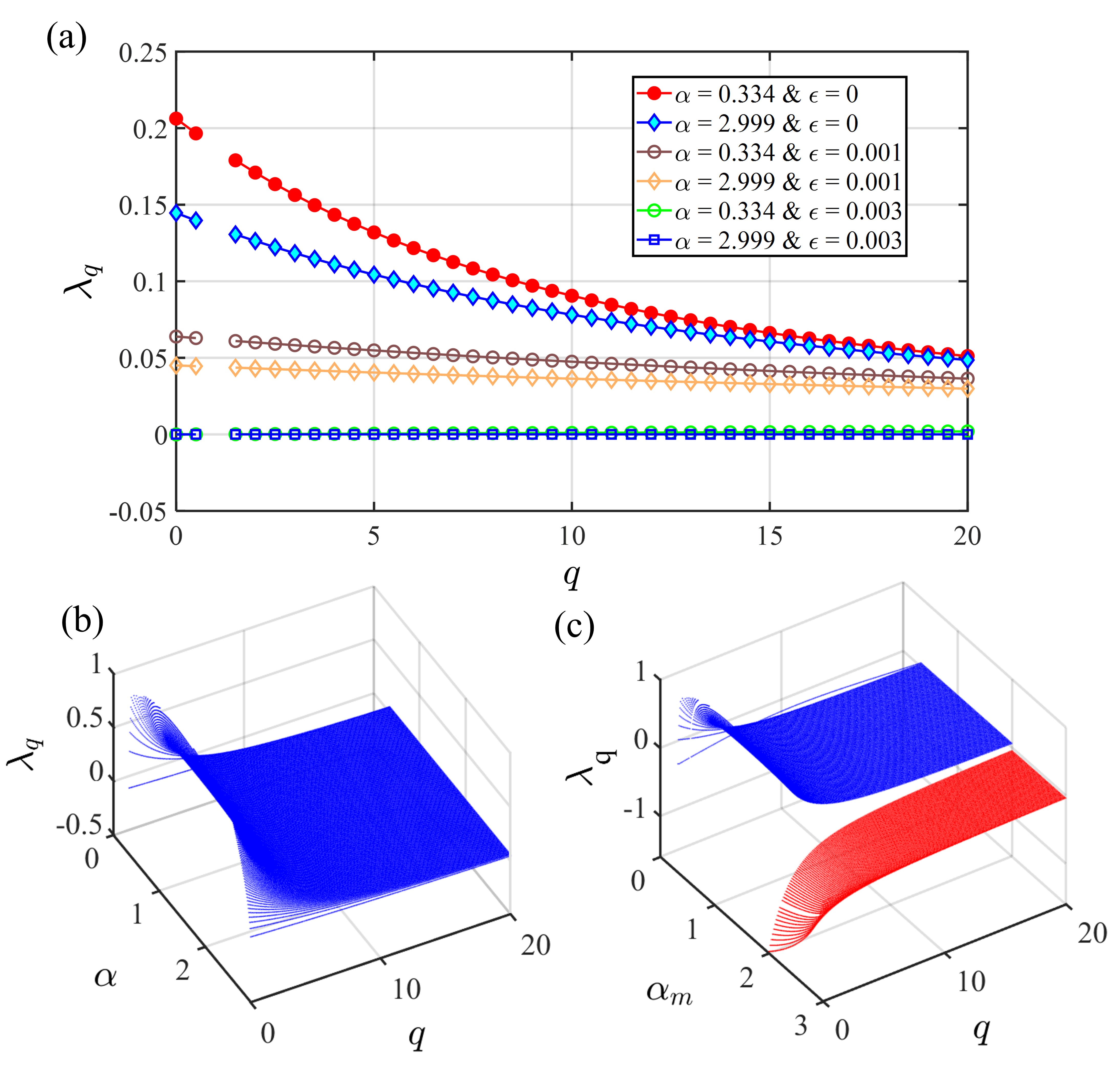}
	\caption{{\bf $q$-generalized Lyapunov exponent $\lambda_q$ illustrating the effect of dynamical control.}
		(a)~$\lambda_q$ as a function of $q$ for representative values of the control parameter $\alpha$, shown for increasing coupling strength $\epsilon$. Larger $\epsilon$ progressively suppresses intermittency.
		(b)~$\lambda_q$ for the uncontrolled map $\Phi_3(x_n)$, where $\alpha$ is a static parameter ($\epsilon=0$).
		(c)~$\lambda_q$ for the controlled map $\Psi_3(x_m,\alpha_m)$, where $\alpha_m$ evolves dynamically according to Eq.~\eqref{eq:PS}. Here, $\alpha$ on the axis denotes the nominal control parameter around which the trajectory $\{\alpha_m\}$ evolves. The coupling strength is fixed at $\epsilon=0.1$. The feedback-driven evolution of $\alpha_m$ suppresses the regions with positive $\lambda_q$ observed in the uncontrolled case, demonstrating the reduction of intermittent chaos.}
	\label{Fig.3lam}
\end{figure}

Figure~\ref{Fig.3lam} illustrates the behavior of the generalized Lyapunov exponent under intermittency control. In Fig.~\ref{Fig.3lam}(a), $\lambda_q$ is shown as a function of $q$ for representative values of $\alpha$ near the boundaries of the chaotic regime. In the absence of control ($\epsilon=0$), $\lambda_q$ remains positive, reflecting strong intermittent chaos. As the coupling strength $\epsilon$ increases, the magnitude of $\lambda_q$ is progressively reduced, indicating the suppression of intermittent bursts and the enhancement of laminar dynamics. Figures~\ref{Fig.3lam}(b) and \ref{Fig.3lam}(c) present the generalized Lyapunov spectrum before and after applying control, respectively. While the uncontrolled system exhibits extended regions with positive $\lambda_q$, these regions collapse after control, demonstrating that regulating intermittency provides an effective route to chaos suppression.

\subsection{Sensitivity to initial conditions}
\label{sec:sensitivity}

The analysis of the intermittency transition is completed through the $q$-generalized sensitivity to initial conditions, which the generalized Lyapunov coefficients determine explicitly. Intermittent dynamics are naturally characterized through their sensitivity to initial conditions, which reflects how nearby trajectories separate in time,
\begin{equation}
	\label{eq:xi}
	\xi_t \equiv
	\lim_{\Delta x(0)\to 0}
	\frac{\Delta x(t)}{\Delta x(0)} .
\end{equation}
For fully developed chaos, where the conventional Lyapunov exponent $\lambda_1$ is positive, trajectory separation grows exponentially, $\xi_t = e^{\lambda_1 t}$. In contrast, near critical points associated with intermittency, the dynamics are dominated by long laminar phases interrupted by irregular bursts, leading to a breakdown of exponential instability \cite{Sec3-4,Sec3-4-1,Sec3-4-2,Sec3-4-3}. In this regime, the sensitivity follows a power-law growth rather than a simple exponential form.

Such behavior is naturally captured within the framework of $q$-generalized dynamics based on the nonextensive Tsallis entropy \cite{Tsallis98,Sec3-5,Sec3-5-1}, where the sensitivity to initial conditions is expressed as
\begin{equation}
	\label{eq:xit}
	\xi_t =
	\exp_q(\lambda_q t)
	\equiv
	\left[1-(q-1)\lambda_q t\right]^{-1/(q-1)} .
\end{equation}
The standard exponential sensitivity is recovered in the limit $q\to1$, while $q\neq1$ describes weak chaos characteristic of intermittent systems. The parameter $q$ is referred to as the entropic index. For $q>1$ ($q<1$) and $\lambda_q<0$ ($\lambda_q>0$), the system is said to be weakly insensitive (sensitive) to initial conditions. In particular, for $q>1$, $\xi_t$ exhibits inverse power-law scaling, reflecting the dominance of laminar motion in phase space. The sensitivity-based approach has been successfully applied to a wide variety of nonlinear maps, including logistic \cite{Sec3-6}, $z$-logistic \cite{Sec3-7}, circle \cite{Tsallis98}, and $z$-circular \cite{Sec3-9} maps.

The $q$-generalized Lyapunov exponent is closely connected to the generalized entropy production rate,
\begin{equation}
	K_q = \lim_{t\to\infty} \frac{S_q(t)-S_q(0)}{t},
\end{equation}
where
\begin{equation}\label{eq:Sq}
	S_q=\frac{1-\sum_{i}^{W}p_i^q}{1-q}
\end{equation}
is the Tsallis entropy, a nonextensive generalization of the conventional Boltzmann--Gibbs entropy $S=-\sum_i p_i \ln p_i$. It has been conjectured that the relation between the asymptotic entropy production rate and dynamical instability---known as the Pesin identity for fully chaotic systems---can be extended to the edge of chaos. For ergodic one-dimensional maps, the Kolmogorov--Sinai entropy equals the Lyapunov exponent, $K_{\mathrm{KS}}=\lambda_1$. For systems exhibiting weak chaos, characterized by power-law sensitivity to initial conditions, a generalized Pesin-like identity has been proposed in the form $K_q=\lambda_q$ for $\lambda_q>0$ \cite{Sec3-6,Sec3_3-2m,Sec3_3-3m,Sec3_3-4m,Sec3_3-5m,Sec3_3-1m}. The validity of this relation is closely connected to the existence of an invariant SRB measure and well-defined $q$-generalized Lyapunov exponents over finite parameter intervals---both of which hold exactly for the maps considered here.

Using the generalized Lyapunov exponents given in Eqs.~\eqref{eq:GLE} and \eqref{eq:GLE2}, together with the $q$-generalized sensitivity of Eq.~\eqref{eq:xit}, we obtain
\begin{widetext}
\begin{equation}\label{eq:xifull}
	\xi_{t}=\exp_q(\lambda_qt)=\left\{
	\begin{array}{ll}
		\left[1-(q-1)\left[\dfrac{\left(\frac{3(1+\beta+2\sqrt{\beta})^2}{(1+3\beta)(3+\beta)}
			\right)^{1-q}-1}{1-q}\right]t\right]^{\frac{-1}{q-1}} & \text{(control OFF)},\\[24pt]
		\Big[1-\big(1-\big[\Gamma(\beta,\epsilon)\big]^{1-q}\big)\,t\Big]^{\frac{1}{1-q}} & \text{(control ON)},
	\end{array}
	\right.
\end{equation}
\end{widetext}
with $\Gamma(\beta,\epsilon)$ given by Eq.~\eqref{eq:GLE32}. Using the invariant measure, we have thus analytically evaluated the $q$-generalized Lyapunov exponents of these maps, which are independent of the initial conditions \cite{Sec3-11,Sec3-12}. The resulting power-law form of the sensitivity to initial conditions provides a natural and effective characterization of intermittent dynamics, where long laminar phases coexist with rare chaotic bursts. In parallel, we performed numerical simulations of the sensitivity to initial conditions and found excellent agreement with the analytical predictions, including the specific values of the entropic index $q$ and the generalized Lyapunov exponent $\lambda_q$ at the intermittency transition. This formulation provides a unified description of intermittent chaotic dynamics and their regulation through dynamical control, naturally interpolating between weakly chaotic and regular regimes.

\section{Application to interacting microbubble clusters}
\label{sec:bubbles}

\begin{figure*}[ht]
	\includegraphics[width=\textwidth]{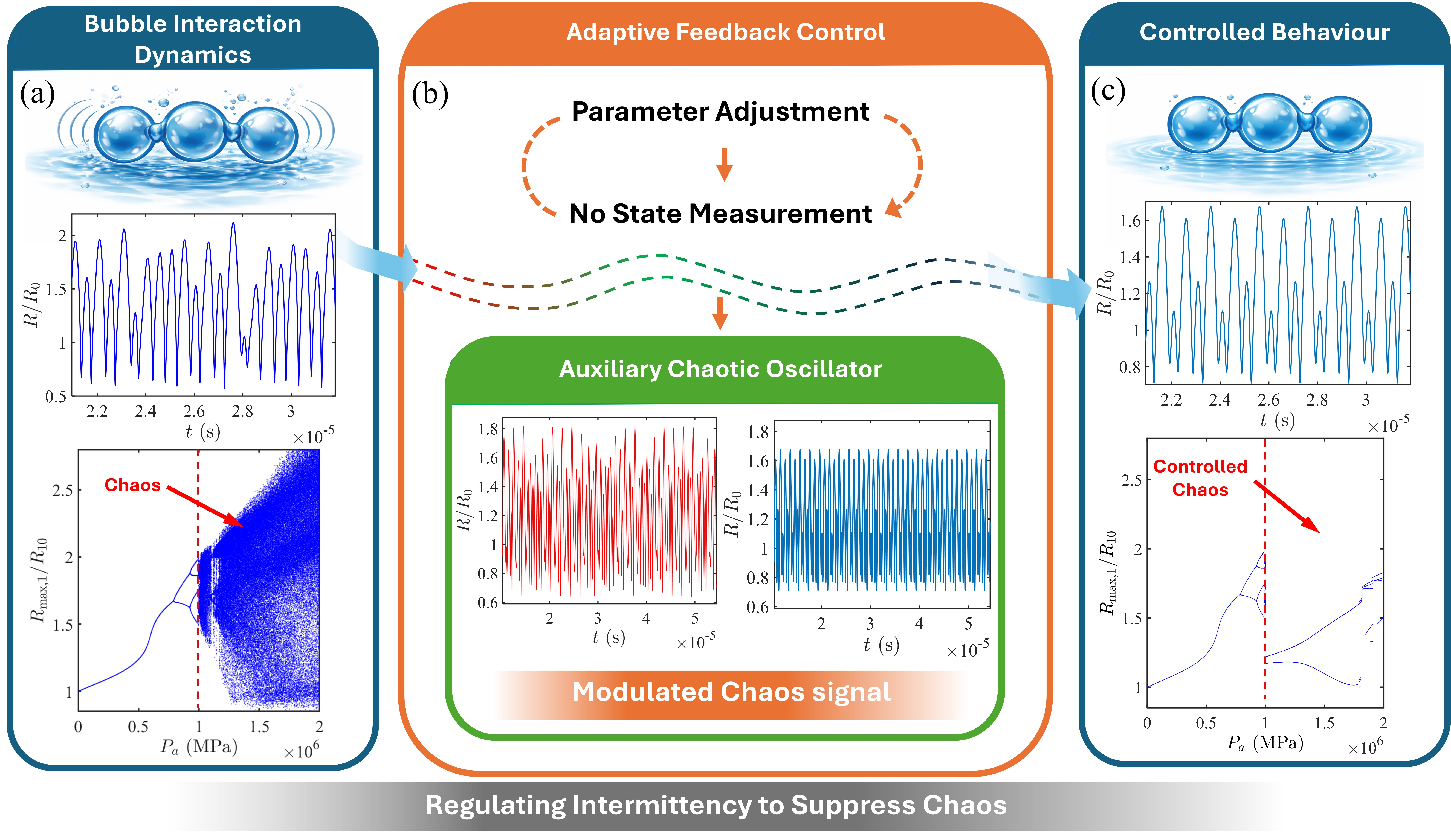}
	\caption{{\bf Overview of intermittency-based chaos control in microbubble dynamics.}
		Numerical results illustrating the adaptive control mechanism applied to a system of three interacting microbubbles described by the Keller--Herring model.
		(a)~Top: schematic representation of chaotic oscillations in a three-bubble system. Middle: time series of the normalized bubble radii for three bubbles with initial radii $R_{10}=4\,\mu$m, $R_{20}=5\,\mu$m, and $R_{30}=6\,\mu$m, driven at a frequency of $2$~MHz and pressure of $1.5$~MPa. Bottom: bifurcation diagram of the normalized three-bubble radius versus driving pressure at fixed frequency $2$~MHz for the same initial radii.
		(b)~Time series before (left), exhibiting intermittent chaotic behavior, and after (right) applying the control method at pressure $1$~MPa and $\epsilon_t=0.005$.
		(c)~Top: schematic representation of stable oscillations in a three-bubble system. Middle and bottom: the corresponding controlled dynamics under the same parameters, where intermittent bursts are suppressed and stable oscillations are recovered; here $\epsilon_t=0.1$.
		Detailed bifurcation diagrams, Lyapunov spectra, time series, and phase-space portraits corresponding to each stage of this figure are presented in Figs.~\ref{Fig.SB1}--\ref{Fig.SB9}.}
	\label{Fig.overview}
\end{figure*}

To assess the physical relevance of the proposed intermittency-control framework, we apply it to a realistic nonlinear system consisting of interacting ultrasound-driven microbubbles. Microbubble clusters are well known to exhibit strong nonlinear responses, including intermittency, multistability, and fully developed chaos, depending sensitively on the acoustic driving conditions. Figure~\ref{Fig.overview} provides an overview of the analysis developed in this section: the uncontrolled cluster exhibits intermittent chaotic radial oscillations and a dense bifurcation structure, while under adaptive frequency modulation the intermittent bursts are suppressed and stable collective oscillations are recovered. The remainder of this section presents the underlying model (Sec.~\ref{sec:KH}), the derivation of the control law for the physically accessible parameter (Sec.~\ref{sec:bubblecontrol}), and the complete set of bifurcation diagrams, Lyapunov spectra, time series, and phase-space portraits before (Sec.~\ref{sec:uncontrolledbubbles}) and after (Sec.~\ref{sec:controlledbubbles}) control, together with all parameter values (Tables~\ref{Table1}--\ref{Table3}). The numerical procedures are detailed in Appendix~\ref{app:numerics}.

\subsection{Theoretical model for a cluster of microbubbles under periodic acoustic forcing}
\label{sec:KH}

The dynamics of a cluster of $N_b$ interacting free gas microbubbles, modeled as ultrasound contrast agents (UCAs) with a thin elastic shell, is described by the generalized Keller--Herring (K--H) equation \cite{I-34,I-8m}
\begin{widetext}
\begin{equation}\label{eq:KH}
	\left[1-(1+\kappa)\frac{\dot{R}_i}{c}\right]R_i\ddot{R}_i+\frac{3}{2}
	\left[1-(3\kappa+1)\frac{\dot{R}_i}{3c}\right]\dot{R}^2_i
	=\frac{1}{\rho}\left[1+(1-\kappa)\frac{\dot{R}_i}{c}+\frac{R_i}{c}\frac{d}{dt}\right]
	\big[P_i(R,\dot{R})-P_{\infty}(t)\big]
	-\sum^{N_b}_{\substack{j=1\\ j\neq i}}\frac{R_j}{d_{ij}}\big(R_j\ddot{R}_j+2\dot{R}^2_j\big),
\end{equation}
\end{widetext}
where $i=1,2,\ldots,N_b$. Here $R_i$ denotes the radius of bubble $i$, $R_{i0}$ its equilibrium radius, while $\dot{R}_i$ and $\ddot{R}_i$ represent the corresponding wall velocity and acceleration, respectively. The parameters $c$ and $\rho$ denote the speed of sound and the density of the surrounding liquid. The pressure in the liquid far from the bubble is given by $P_{\infty}(t)=P_0+P_{ac}(t)$, where $P_0$ is the ambient static pressure and $P_{ac}(t)$ represents the externally applied acoustic forcing, taken as $P_{ac}(t)=P_a \sin(2\pi f t)$. Throughout this section we denote the K--H interpolation parameter by $\kappa$, the shell thickness by $\delta$, and the polytropic exponent by $\gamma_p$, to avoid any confusion with the map parameters $\beta$ and $\epsilon$ and the function $\Gamma$ of the preceding sections.

In addition to the external acoustic field, each oscillating bubble generates a secondary pressure that acts on its neighboring bubbles. This mutual interaction is accounted for by the final term on the right-hand side of Eq.~\eqref{eq:KH}. The governing equation is referred to as the K--H equation, as it continuously interpolates between the Keller and Herring models depending on the value of the dimensionless parameter $\kappa$. Specifically, $\kappa=0$ and $\kappa=1$ recover the Keller-type and Herring-type formulations, respectively. Prosperetti and Lezzi \cite{I-8m} have shown that both models exhibit qualitatively similar dynamical behavior. It is also worth noting that, in the limit $c\rightarrow\infty$, the liquid may be regarded as incompressible, in which case Eq.~\eqref{eq:KH} reduces to the classical Rayleigh--Plesset--Noltingk--Neppiras--Poritsky (RPNNP) equation for bubble dynamics in an incompressible fluid \cite{I-8m}.

To incorporate the effect of the thin elastic shell surrounding the UCA microbubbles, an explicit expression for the interfacial pressure $P_i(R,\dot{R})$ must be introduced, which requires several additional material parameters in the general K--H model. The presence of the encapsulating shell gives rise to pronounced nonlinear acoustic responses; such nonlinearities form the physical basis of advanced ultrasonic imaging techniques, including harmonic and pulse-inversion imaging \cite{I-35m}. Following Morgan {\it et al.} \cite{I-36}, the interfacial pressure is given by
\begin{multline}\label{eq:Pi}
	P_i(R,\dot{R})=\left(P_{0}
	+\frac{2(\sigma+\chi)}{R_{i0}}\right)\left(\frac{R_{i0}}{R_{i}}\right)^{\!3\gamma_p}
	-\frac{4\mu\dot{R}_{i}}{R_{i}}-\frac{2\sigma}{R_{i0}}\\
	-\frac{2\chi}{R_{i}}\left(\frac{R_{i0}}{R_{i}}\right)^{\!2}
	-12\mu_{sh}\delta\,\frac{\dot{R}_{i}}{R_{i}(R_{i}-\delta)}.
\end{multline}
Here $P_i(R,\dot{R})$ denotes the pressure on the liquid side of the bubble--liquid interface for bubble $i$. The parameter $\mu_{sh}$ is the viscosity of the encapsulating shell, $\delta$ is the shell thickness, $\mu$ is the viscosity of the surrounding liquid, $\gamma_p$ is the dimensionless polytropic exponent of the gas inside the UCA, $\chi$ represents the shell elasticity, and $\sigma$ is the surface tension. The model has previously been solved for isolated microbubbles using the set of physical parameters listed in Table~\ref{Table1} \cite{I-9,I-10,I-32}. The same parameter values are adopted throughout the present study to ensure consistency with established experimental and theoretical investigations.

\begin{figure*}
	\begin{center}
		\begin{minipage}{0.49\textwidth}\centering (a)\\ \includegraphics[width=\linewidth]{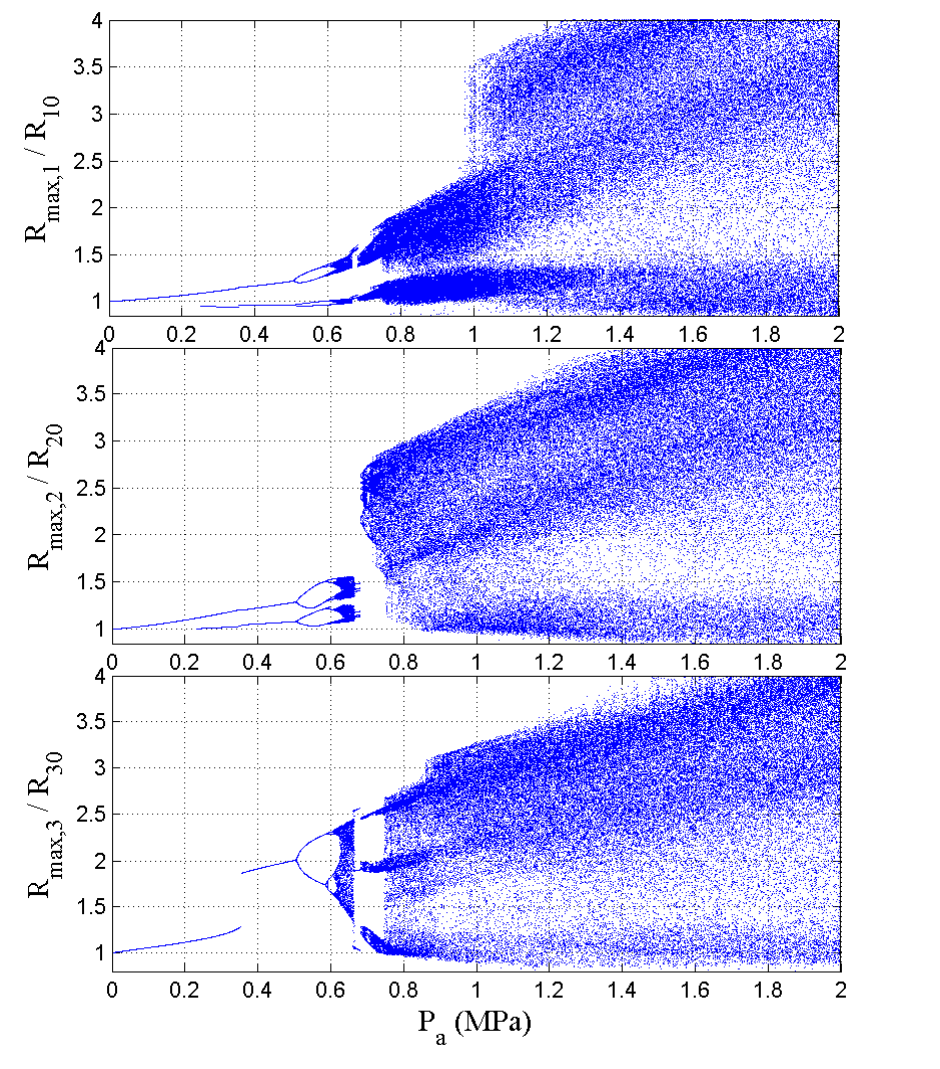}\end{minipage}
		\begin{minipage}{0.49\textwidth}\centering (b)\\ \includegraphics[width=\linewidth]{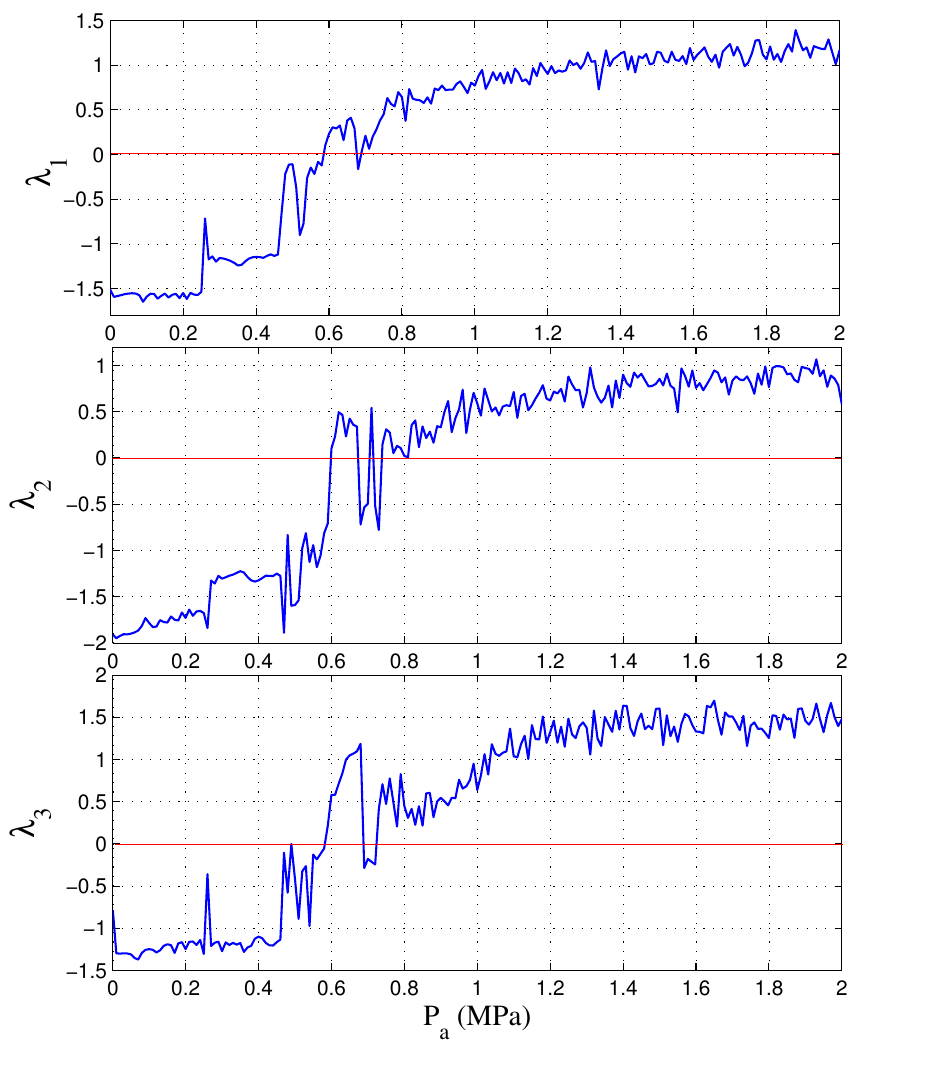}\end{minipage}
		\caption{(a)~Bifurcation diagrams and (b)~the corresponding Lyapunov spectra of the normalized three-bubble radii as functions of the driving pressure for $R_{10}=4~\mu$m, $R_{20}=5~\mu$m, $R_{30}=6~\mu$m, and acoustic frequency $f=1$~MHz.}\label{Fig.SB1}
	\end{center}
\end{figure*}

\begin{table}[htbp]
	\begin{center}
		\caption{Constant parameters used in the general Keller--Herring equation for an ultrasound contrast agent microbubble (for a bubble/water system at $20^\circ$C).}
		\label{Table1}
		\begin{tabular}{lccc}
			\hline\noalign{\smallskip}
			Symbol & Description &Units&Value \\
			\noalign{\smallskip}\hline\noalign{\smallskip}
			$\mu$    &   Liquid viscosity        & $ \textmd{Ns/m}^2$    & $0.001$ \\
			$\sigma$ &   Surface tension  &  $ \textmd{N/m}$      &$0.072$ \\
			$c$      &   Sound velocity   & $\textmd{m/s}$        & $1481$ \\
			$P_0$    &   Static ambient pressure   &  $\textmd{N/m}^2$ & $1.01 \times 10^5$ \\
			$\rho$   &   Liquid density   &$\textmd{kg/m}^3$ &$998$ \\
			$\chi$   & Shell elasticity   &  $\textmd{N/m}$ &$8$ \\
			$\delta$& Shell thickness   &  $\textmd{m}$  & $15\times10^{-9}$  \\
			$\mu_{sh}$& Shell viscosity        &    $\textmd{Pa}\;\textmd{s}$      &                $1.77$\\
			$\kappa$& K--H parameter (Keller type)   & --- &$0$  \\
			$\gamma_p$& Polytropic exponent  & --- &$1.33$  \\
			$d_{12}$& Separation, bubbles 1--2  &$\textmd{m}$& $100\times10^{-6}$\\
			$d_{13}$& Separation, bubbles 1--3  &$\textmd{m}$& $150\times10^{-6}$\\
			$d_{23}$& Separation, bubbles 2--3  &$\textmd{m}$& $200\times10^{-6}$\\
			\noalign{\smallskip}
			\hline
		\end{tabular}
	\end{center}
\end{table}

We assume that the interbubble distance $d_{ij}$ between the centers of bubbles $i$ and $j$ is sufficiently large to ensure that the bubbles remain spherical throughout their motion. Following Mettin {\it et al.} \cite{I-37}, the time delay $\tau = d_{ij}/c$ associated with the propagation of acoustic disturbances between neighboring bubbles is assumed to be negligible. This delay arises from the finite time required for pressure waves to travel through the surrounding liquid over the distance $d_{ij}$. As demonstrated in the models of Takahira {\it et al.} \cite{I-34} and Macdonald {\it et al.} \cite{I-32}, when the equilibrium bubble radius is of the order of $R_{i0}\sim10~\mu\mathrm{m}$ and the interbubble distance is of the order of $d_{ij}\sim100~\mu\mathrm{m}$, the ratio of the delay time to the natural oscillation period of the bubbles is approximately $2\%$, rendering its effect negligible. In the present work, we choose $d_{ij}$ to be much larger than the equilibrium radius, specifically $d_{ij}\approx30R_{i0}$. By comparing the maximum delay time $\tau$ with the oscillation period $T=1/f$, we obtain $\tau<T$, confirming that propagation-delay effects do not play a significant role in the system dynamics. Consequently, the compressibility effects of the liquid can be neglected, and the fluid is treated as effectively incompressible throughout the analysis.

\begin{figure*}
	\begin{center}
		\begin{minipage}{0.49\textwidth}\centering (a)\\ \includegraphics[width=\linewidth]{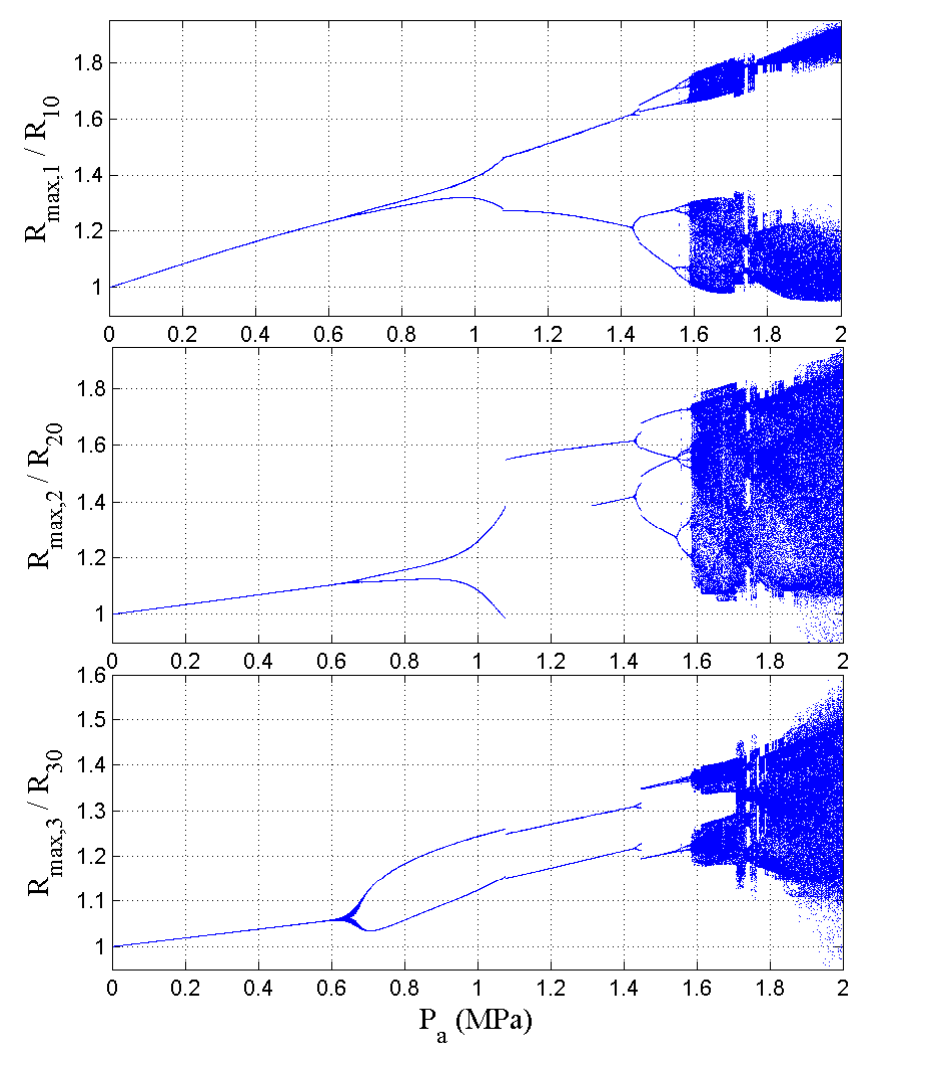}\end{minipage}
		\begin{minipage}{0.49\textwidth}\centering (b)\\ \includegraphics[width=\linewidth]{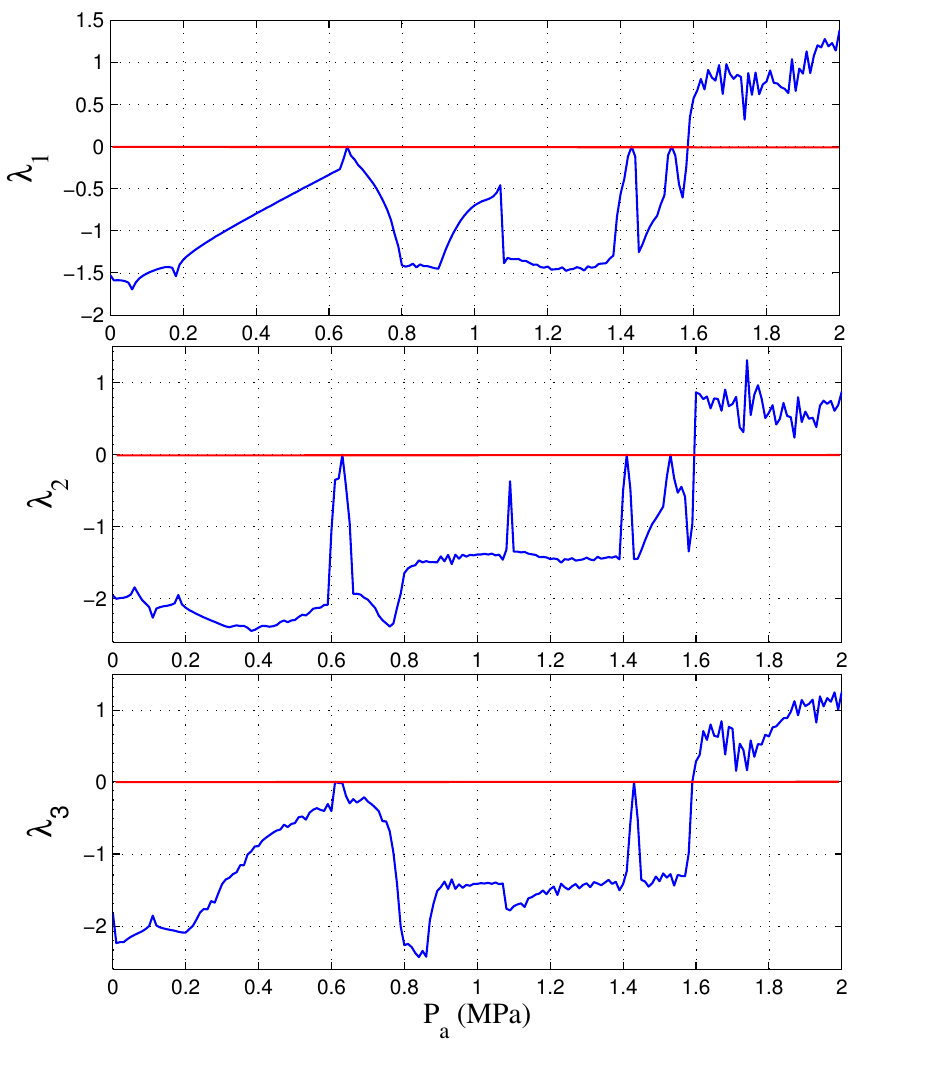}\end{minipage}
		\caption{(a)~Bifurcation diagrams and (b)~the corresponding Lyapunov spectra of the normalized radii of a three-microbubble cluster as functions of the driving pressure for $R_{10}=4~\mu$m, $R_{20}=5~\mu$m, $R_{30}=6~\mu$m, and acoustic frequency $f=3$~MHz.}\label{Fig.SB2}
	\end{center}
\end{figure*}

For the purpose of stability analysis, it is convenient to rewrite the second-order differential equation~\eqref{eq:KH} as an autonomous system of first-order differential equations. Applying the coupled nonlinear equation~\eqref{eq:KH} to each microbubble in the cluster and performing an order-reduction procedure yields a system of $2N_b$ first-order differential equations of the compact form
\begin{equation}\label{eq:compactflow}
	J\,\frac{dV}{dt}=F(V,\bm{p}),
\end{equation}
where $V=(R_1,\ldots,R_i,\dot{R}_1,\ldots,\dot{R}_i,\theta)$ denotes an autonomous vector field, $\theta = f t$ is a cyclic variable, and the parameter vector
\begin{equation}
	\bm{p} = \big(R_{i0}, P_0, P_a, f, \mu, \mu_{sh}, \kappa, \rho, c, \gamma_p, \delta, \chi, \sigma\big)
\end{equation}
belongs to the associated parameter space. The explicit matrix form of Eq.~\eqref{eq:compactflow}, together with the functions $\mathcal{A}_i(\vec{x})$ and $\mathcal{B}(R_i,\dot{R}_i,R_{i0})$ entering it, is given in Appendix~\ref{app:matrix}.
The resulting dynamical system generates a flow $\Phi = \{\Phi^{T}\}$ on the phase space $M = \mathbb{R}^{2N_b} \times S^{1}$, and there exists an associated global (Poincar\'e) map
\begin{equation}
	P : \Sigma_c \longrightarrow \Sigma_c,\qquad
	V_P \longmapsto P(V_P)=\{\Phi^T\}\big|_{\Sigma_c},
\end{equation}
where $T = 1/f$ denotes the acoustic period, $\theta_0$ is a constant specifying the Poincar\'e cross-section, and
\begin{equation}
	\Sigma_c=\big\{(R_1,\ldots,R_i,\dot{R}_1,\ldots,\dot{R}_i,\theta)\in \mathbb{R}^{2N_b}\times S^{1} \,\big|\, \theta=\theta_0\big\}.
\end{equation}
The choice of the Poincar\'e section is arbitrary; the only requirement is that the system trajectory intersects the section once during each acoustic cycle. For driven oscillators such as the present bubble model, a natural way to define the Poincar\'e section is to cut the torus-like phase space transversely to the cyclic $\theta$ direction at a fixed value $\theta_0$ \cite{S-6}.

\subsection{Choice of the control parameter and adaptive frequency-modulation law}
\label{sec:bubblecontrol}

The interaction among oscillating microbubbles is highly nonlinear and can give rise to strongly chaotic dynamics. In practical applications, it is therefore essential to suppress unwanted chaotic oscillations in order to optimize the collective response of a bubble cluster. Identifying the physical parameters that govern the onset and intensity of chaotic oscillations is crucial for designing appropriate insonation conditions and achieving controlled microbubble behavior in clinical applications. Several control strategies have been demonstrated to be effective in regulating chaotic oscillations of microbubbles, including dual-frequency acoustic forcing achieved by applying a periodic perturbation \cite{Behnia-C}, and variation of bubble-cluster size, accounting for interbubble coupling and bubble-size effects \cite{I-31}.

More broadly, practical approaches to chaos control \cite{Sec4-10m,Sec4-11m,Sec4-12m} can be classified into feedback methods (identification of stable and unstable manifolds in the Poincar\'e section; feedback-based control procedures) and non-feedback methods (small modulation of a system control parameter; imposition of a prescribed target dynamics) \cite{Sec4-10m,S-4}. Small-parameter modulation methods, in particular, have been extensively investigated for suppressing chaotic behavior in a variety of dynamical models. However, these techniques have not been shown to be universally effective \cite{S-5}, as they may introduce additional chaotic dynamics arising from error signals generated by discrepancies between the instantaneous system response and its delayed values.

\begin{figure*}
	\begin{center}
		\begin{minipage}{0.49\textwidth}\centering (a)\\ \includegraphics[width=\linewidth]{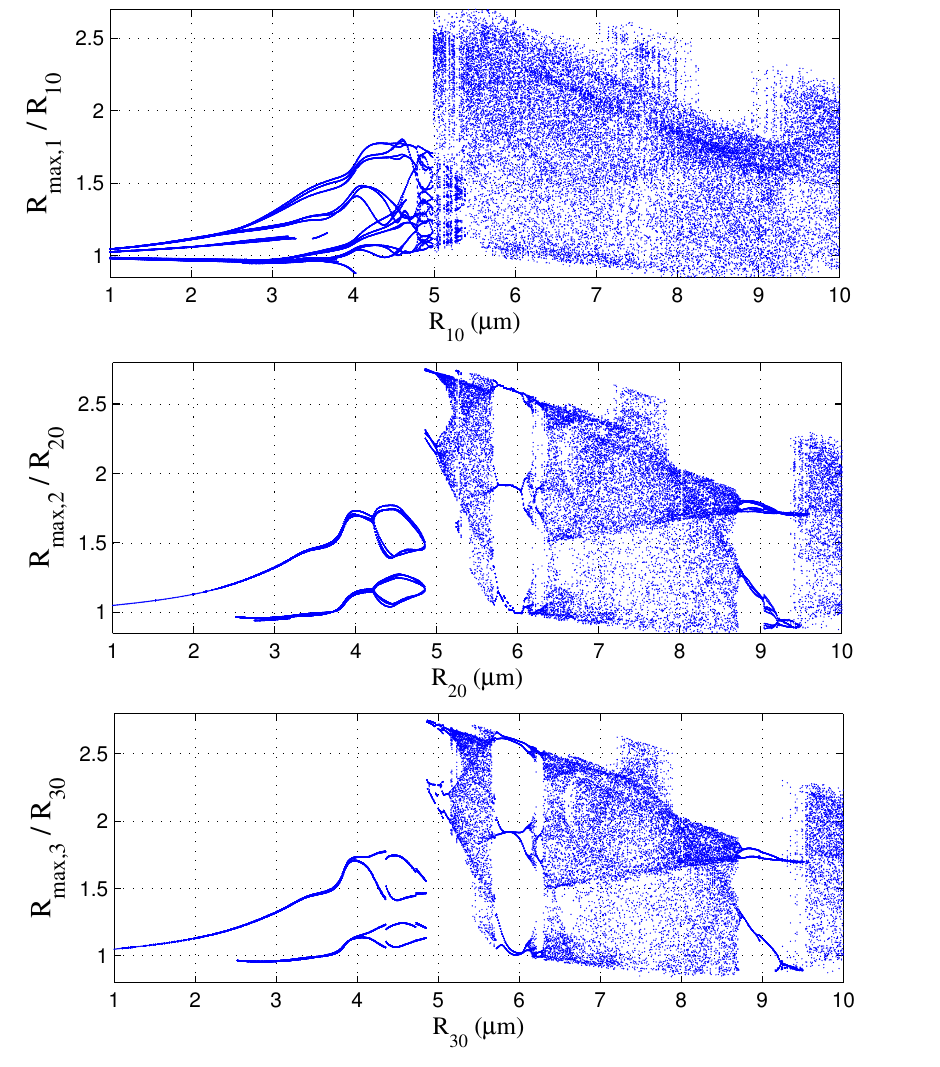}\end{minipage}
		\begin{minipage}{0.49\textwidth}\centering (b)\\ \includegraphics[width=\linewidth]{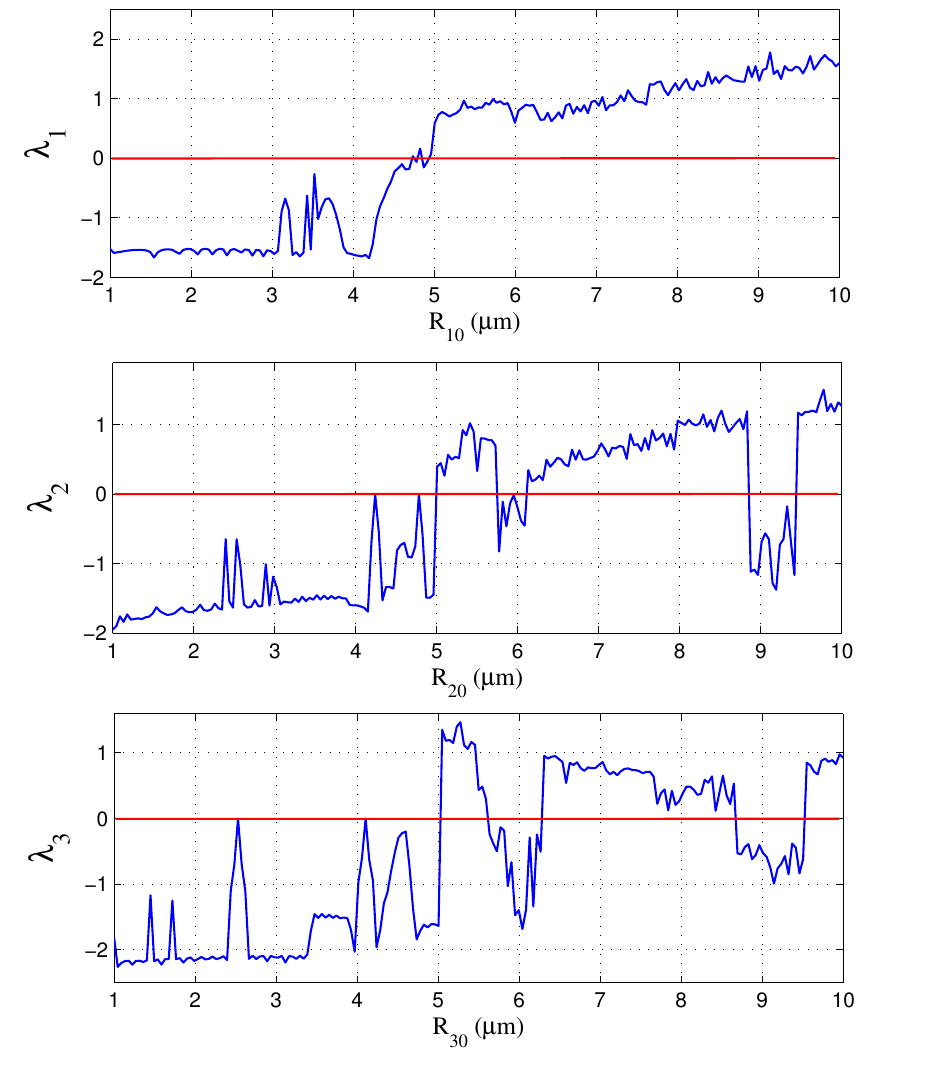}\end{minipage}
		\caption{(a)~Bifurcation diagrams and (b)~the corresponding Lyapunov spectrum of the normalized three-bubble radius as a function of the initial bubble radius, for a driving frequency of $1$~MHz and an acoustic pressure of $700$~kPa.}\label{Fig.SB3}
	\end{center}
\end{figure*}

Before assessing possible chaos-control strategies for a microbubble cluster, it is necessary to identify which parameters of the K--H model can be practically accessed by a control mechanism. Among the parameters entering the Keller--Herring formulation, most material and geometrical quantities are fixed by the bubble composition and surrounding medium and cannot be varied dynamically in practice. In realistic applications involving cavitation phenomena, direct manipulation of internal system parameters is not feasible; control can be applied only through the external acoustic forcing term $P_{ac}(t)$. In practice, the acoustic pressure amplitude and driving frequency are dictated by the specific clinical or technological application. Although acoustic-frequency modulation has been widely employed in studies of cavitation and microbubble dynamics, systematic investigations of the effectiveness of feedback-based control schemes within microbubble cluster systems remain limited. Motivated by the demonstrated effectiveness of feedback-based control approaches, their relative simplicity of implementation, and the practical feasibility of acoustic parameter modulation, we select the driving frequency $f$ as the control parameter and allow it to evolve dynamically according to an auxiliary nonlinear process. This choice enables a direct physical realization of intermittency regulation through frequency modulation, providing a natural bridge between the abstract control scheme developed for one-dimensional maps and experimentally relevant microbubble dynamics.

\begin{figure*}
	\begin{center}
		\begin{minipage}{0.49\textwidth}\centering (a)\\ \includegraphics[width=\linewidth]{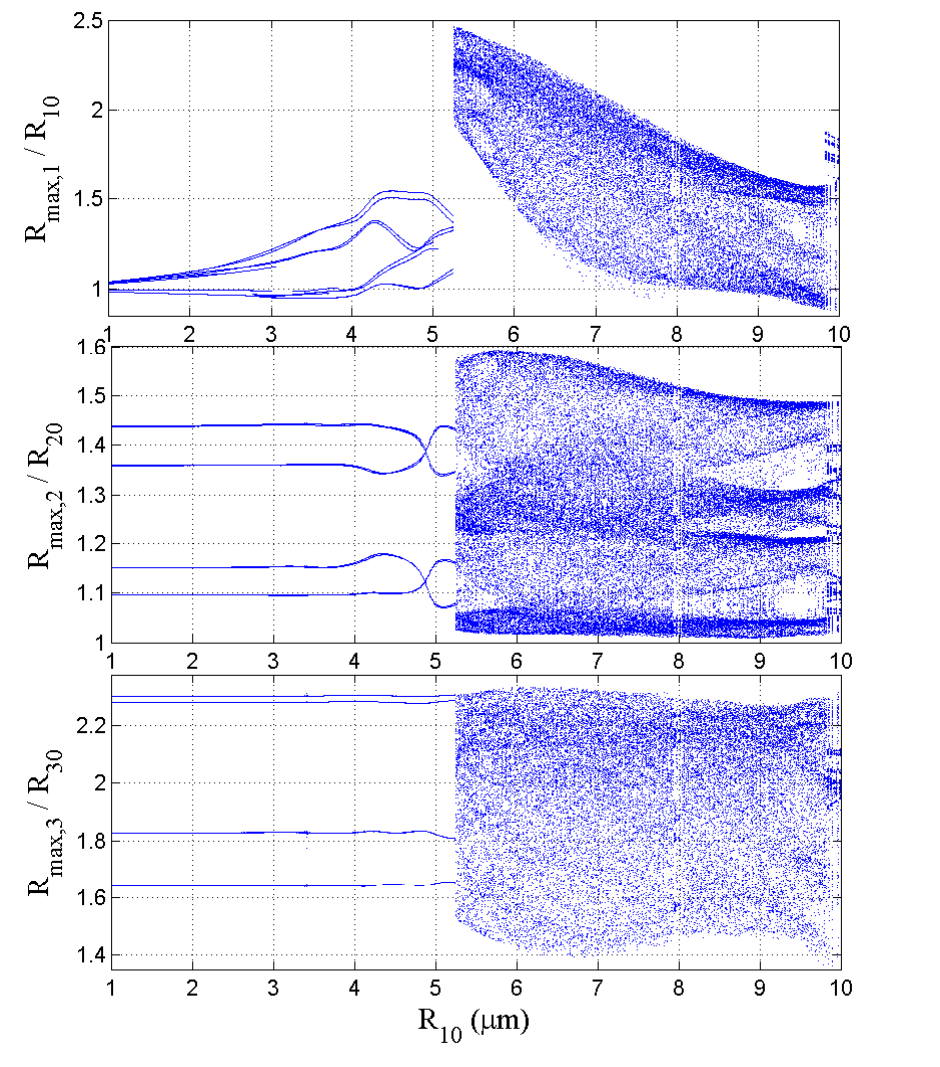}\end{minipage}
		\begin{minipage}{0.49\textwidth}\centering (b)\\ \includegraphics[width=\linewidth]{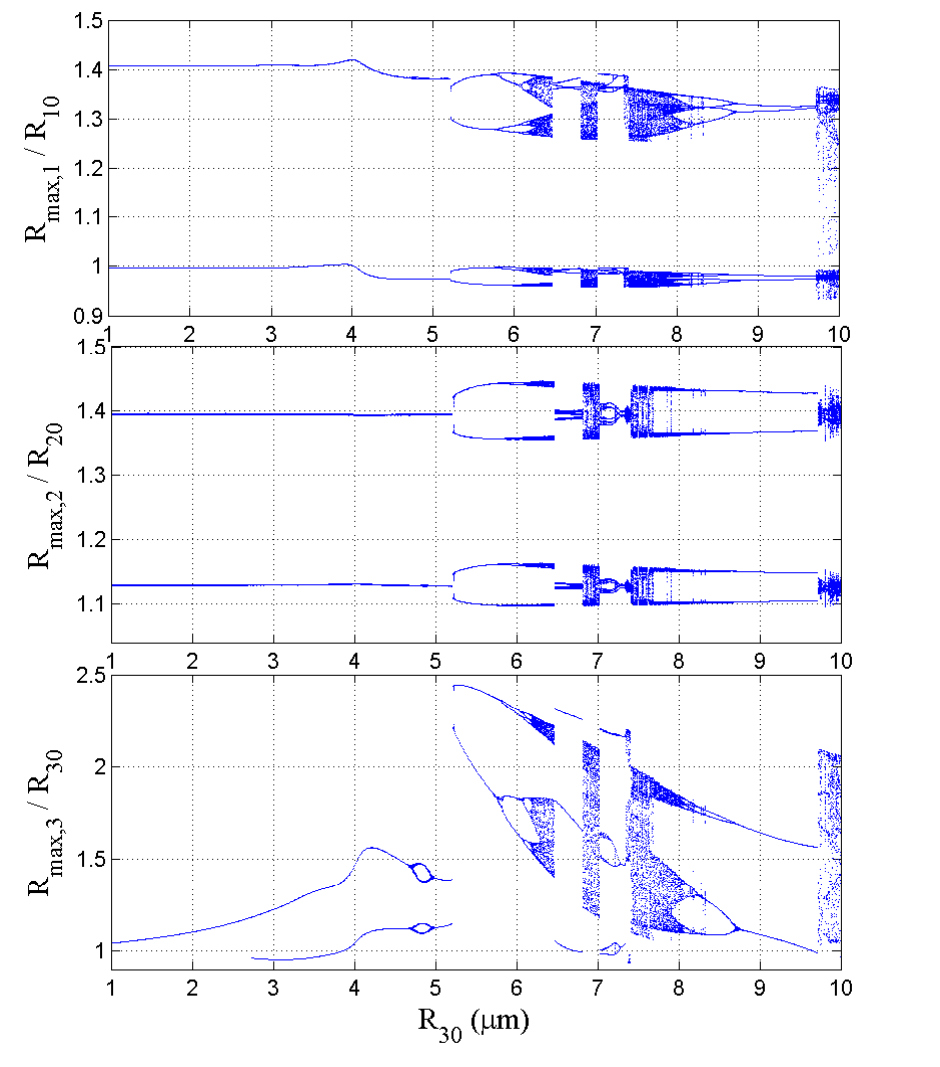}\end{minipage}
		\caption{Bifurcation diagrams of the normalized three-bubble radius as a function of (a)~the initial radius $R_{10}$ at a driving frequency of $1$~MHz and an acoustic pressure of $600$~kPa ($R_{20}=5~\mu\mathrm{m}$, $R_{30}=6~\mu\mathrm{m}$), and (b)~the initial radius $R_{30}$ under the same driving conditions ($R_{10}=4~\mu\mathrm{m}$, $R_{20}=5~\mu\mathrm{m}$).}\label{Fig.SB4}
	\end{center}
\end{figure*}

The frequency-modulation law is not introduced phenomenologically, but follows from the same conjugate-map structure developed in Sec.~\ref{sec:control}. We make use of the conjugate map for $N=2$ [Eq.~\eqref{eq:S7}] and identify the map variable with the discrete-time sequence of the driving frequency, $x_m \equiv f_m$, so that the nonlinear structure of the conjugate map directly governs the evolution of $f$. Within the adaptive control framework of Sec.~\ref{sec:construction}, the static parameter $\alpha$ is replaced by the effective dynamical parameter $g(\alpha_m)$, and the conjugate map can be written as
\begin{equation}
	\tilde{\Phi}_{2}(f_m)
	=
	\frac{1}{g^2(\alpha_m)}\frac{4f_m}{(1-f_m)^2},
\end{equation}
such that
\begin{equation}
	g^2(\alpha_m)\,\tilde{\Phi}_{2}(f_m)
	=
	\frac{4f_m}{(1-f_m)^2}.
\end{equation}
Rather than iterating this map directly, we construct a continuous-time evolution that preserves its nonlinear structure. Following the adaptive framework, we introduce a map-induced flow by defining the evolution as the deviation of the nonlinear transformation from the identity,
\begin{equation}
		\begin{split}
	f_{m+1}
	=&
	f_m
	+
	\frac{\Delta t}{\epsilon_t}
	\left[
	g^2(\alpha_m)\,\tilde{\Phi}_{2}(f_m)-1
	\right]\\
	=&
	f_m
	+
	\frac{\Delta t}{\epsilon_t}
	\left[
	\frac{4f_m}{(1-f_m)^2}
	-
	1
	\right].
		\end{split}
\end{equation}
Taking the continuum limit $\Delta t \to 0$, we obtain
\begin{equation}\label{eq:Dynamical}
	\dot{f}
	=
	\frac{1}{\epsilon_t}
	\left[
	\frac{4f}{(1-f)^2}
	-
	1
	\right]
	=
	\frac{4 f-(1-f)^2}{\epsilon_t(1-f)^2}.
\end{equation}
This construction shows that the auxiliary dynamics of the driving frequency $f$ follows directly from the conjugate-map structure defined in Sec.~\ref{sec:hierarchy}. The singular denominator $(1-f)^2$ reflects the intrinsic geometry of the map, while the shifted formulation ensures a well-defined flow with nontrivial stationary points: the fixed points of Eq.~\eqref{eq:Dynamical} satisfy $4f=(1-f)^2$, i.e., $f_{\pm}=3\pm2\sqrt{2}$ (in the dimensionless units of the map variable), and the flow drives $f(t)$ toward the vicinity of the singular boundary $f=1$ separating them, which corresponds to a critical bifurcation boundary of the driven bubble system. The parameter $\epsilon_t\in(0,1)$ acts as a design parameter that sets the time scale of the control dynamics, and the adaptive function $g(\alpha_m)$ synchronizes the evolution of the map-induced frequency $f$ with the control trajectory $\alpha_m$, enabling the system to evolve toward stable oscillatory regimes while retaining access to intermittent dynamics.

\begin{figure*}
	\begin{center}
		\begin{minipage}{0.49\textwidth}\centering (a)\\ \includegraphics[width=\linewidth]{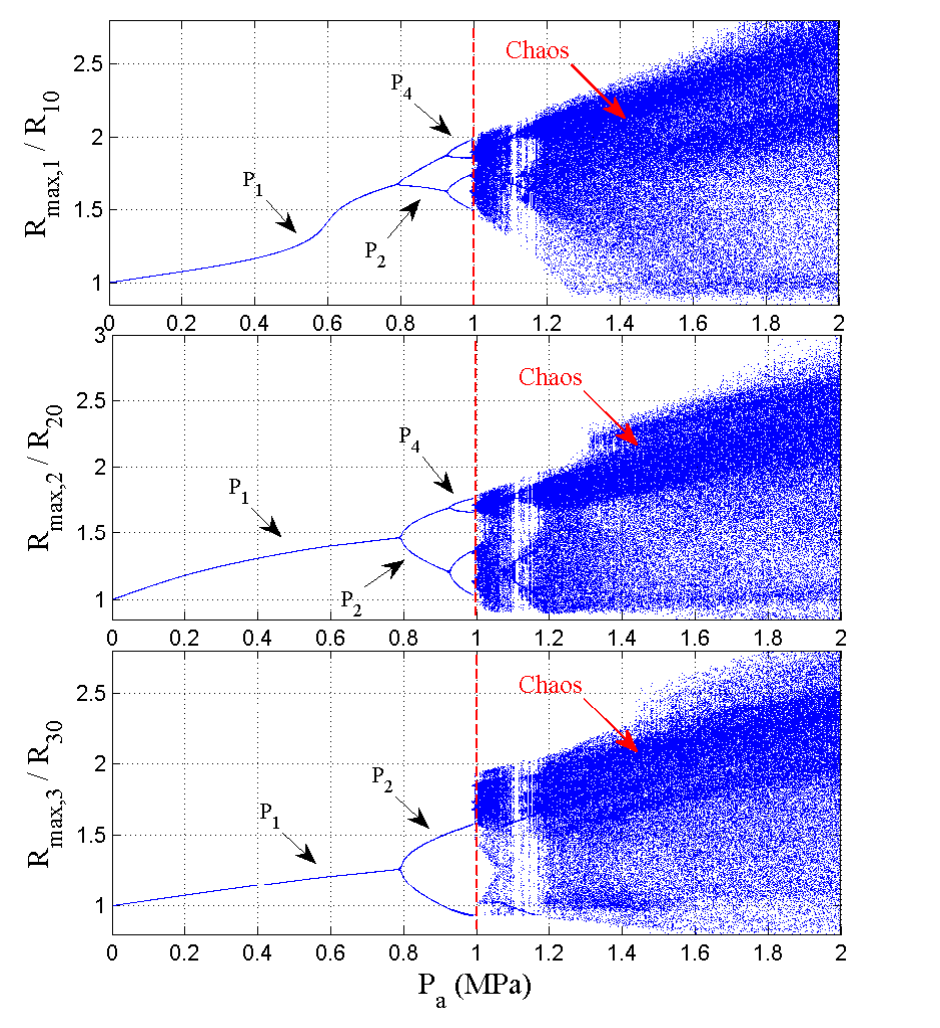}\end{minipage}
		\begin{minipage}{0.49\textwidth}\centering (b)\\ \includegraphics[width=\linewidth]{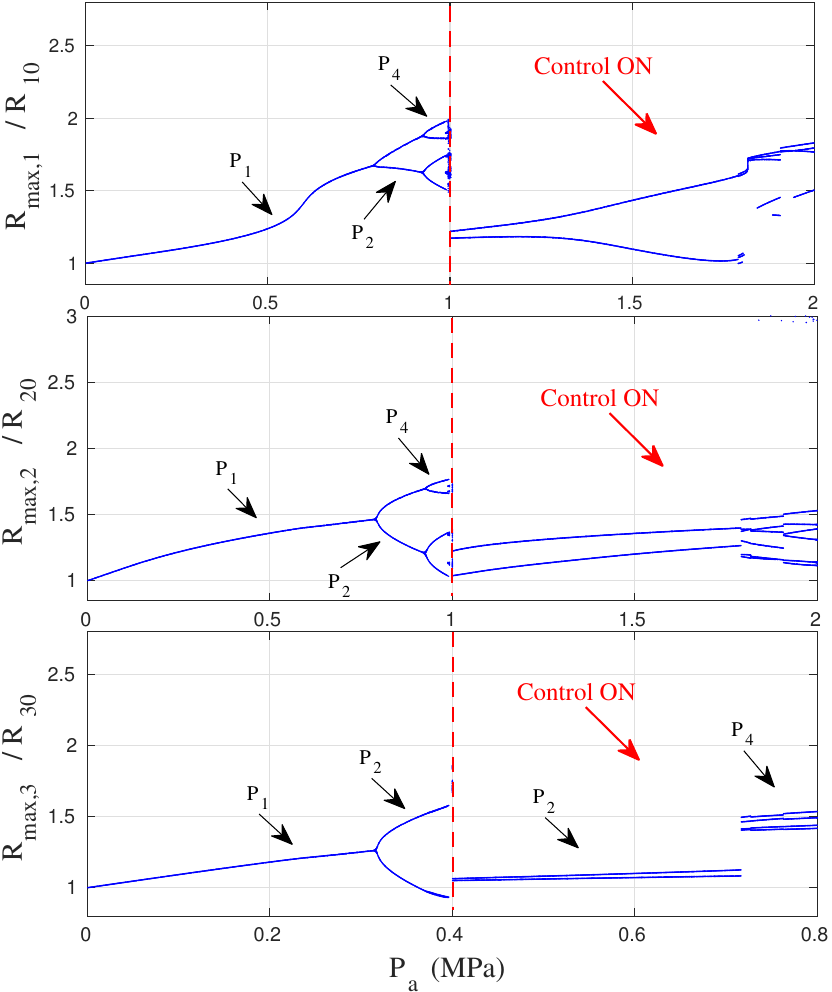}\end{minipage}
		\caption{Bifurcation diagrams of the normalized three-bubble radius driven by a $2$~MHz acoustic field with initial radii $R_{10}=4~\mu\mathrm{m}$, $R_{20}=5~\mu\mathrm{m}$, and $R_{30}=6~\mu\mathrm{m}$ as a function of pressure: (a)~chaotic behavior before applying the proposed control technique, and (b)~stabilized dynamics after control.}\label{Fig.SB5}
	\end{center}
\end{figure*}

Within this framework, the auxiliary system described by Eq.~\eqref{eq:Dynamical} acts as a control-signal generator, while the UCA microbubble dynamics constitutes the controlled system. By considering the UCA microbubble dynamics together with the control system as a coupled $(2N_b+1)$-dimensional dynamical system, a simplified model for stabilizing the radial oscillations of the microbubbles is obtained by augmenting Eq.~\eqref{eq:compactflow} with the additional row given by Eq.~\eqref{eq:Dynamical}, or equivalently
\begin{equation}\label{eq:control-Eq}
	\tilde{J}\frac{d\tilde{V}}{dt}
	=
	\left\{
	\begin{array}{ll}
		J\dfrac{dV}{dt} = F(V,\bm{p}),
		& \text{controlled system}, \\[8pt]
		\text{Eq.~\eqref{eq:Dynamical}},
		& \text{auxiliary control dynamics},
	\end{array}
	\right.
\end{equation}
where $\tilde{V}=(R_1,\ldots,R_i,\dot{R}_1,\ldots,\dot{R}_i,f,\theta)$ represents the augmented autonomous vector field. Within this feedback control scheme, the microbubble system governed by Eq.~\eqref{eq:KH} is driven by the dynamical acoustic pressure
\begin{equation}\label{eq:AC}
	P_{dac}(t) = P_{a}\sin\!\big(2\pi f(t)\, t\big),
\end{equation}
where $P_{dac}(t)$ denotes the dynamical acoustic forcing term and the instantaneous frequency $f(t)$ evolves according to Eq.~\eqref{eq:Dynamical}. As a result, the external forcing continuously adapts to the system dynamics, inducing controlled switching between laminar and chaotic regimes. By embedding the control directly into the acoustic excitation, the feedback mechanism operates intrinsically on the bubble oscillations, enabling intermittent bursts to be progressively suppressed and replaced by stable periodic radial motion. Importantly, this control strategy does not require real-time measurement of the bubble state, but instead relies on a prescribed, globally stable evolution of the driving frequency. Since this control strategy is independent of geometrical considerations, it can be readily extended to high-dimensional dynamical systems, enabling a systematic investigation of interbubble interactions within a controlled framework. From a practical standpoint, suppressing chaos via feedback-based control is particularly advantageous for medical and biomedical applications involving cavitation microbubbles, where stable and predictable oscillatory behavior is essential.

\subsection{Uncontrolled dynamics of a three-bubble cluster}
\label{sec:uncontrolledbubbles}

We first analyze the dynamics of microbubble clusters in an ultrasonic field using standard methods of nonlinear dynamics and deterministic chaos theory (see Appendix~\ref{app:numerics} for the numerical procedures). The study is based on a theoretical description of the microscopic radial motion of individual microbubbles within the cluster. In particular, we consider a system of three interacting microbubbles ($N_b=3$), with their centers located at the vertices of a triangle of varying geometry, and investigate the stability of the interacting microbubbles with particular emphasis on the roles of the equilibrium bubble radius and the acoustic driving pressure amplitude. The primary objective is to characterize the dynamical response of the cluster through bifurcation analysis and evaluation of the maximum Lyapunov exponent, which together provide a comprehensive description of the system's nonlinear behavior.

Figure~\ref{Fig.SB1}(a) presents the pressure--bifurcation diagram for three interacting microbubbles driven at a frequency of $1$~MHz. A key indicator of dynamical instability and the onset of chaos is the maximum Lyapunov exponent; accordingly, Fig.~\ref{Fig.SB1}(b) displays the corresponding Lyapunov spectrum, allowing direct comparison with the bifurcation structure. Two distinct dynamical regimes can be identified. For driving pressures in the range $0 < P_a < 0.6$~MPa, the system remains in a bifurcation regime characterized by negative maximum Lyapunov exponents, indicating stable or periodic oscillations. In contrast, for $P_a > 0.6$~MPa, the maximum Lyapunov exponent becomes positive, signaling the emergence of chaotic oscillations. These results clearly demonstrate that increasing the acoustic pressure amplitude reduces the dynamical stability of the microbubble cluster and promotes chaotic behavior, consistent with previous theoretical and experimental studies of microbubble dynamics \cite{I-32}.

\begin{figure}
	\begin{center}
		\includegraphics[width=\linewidth]{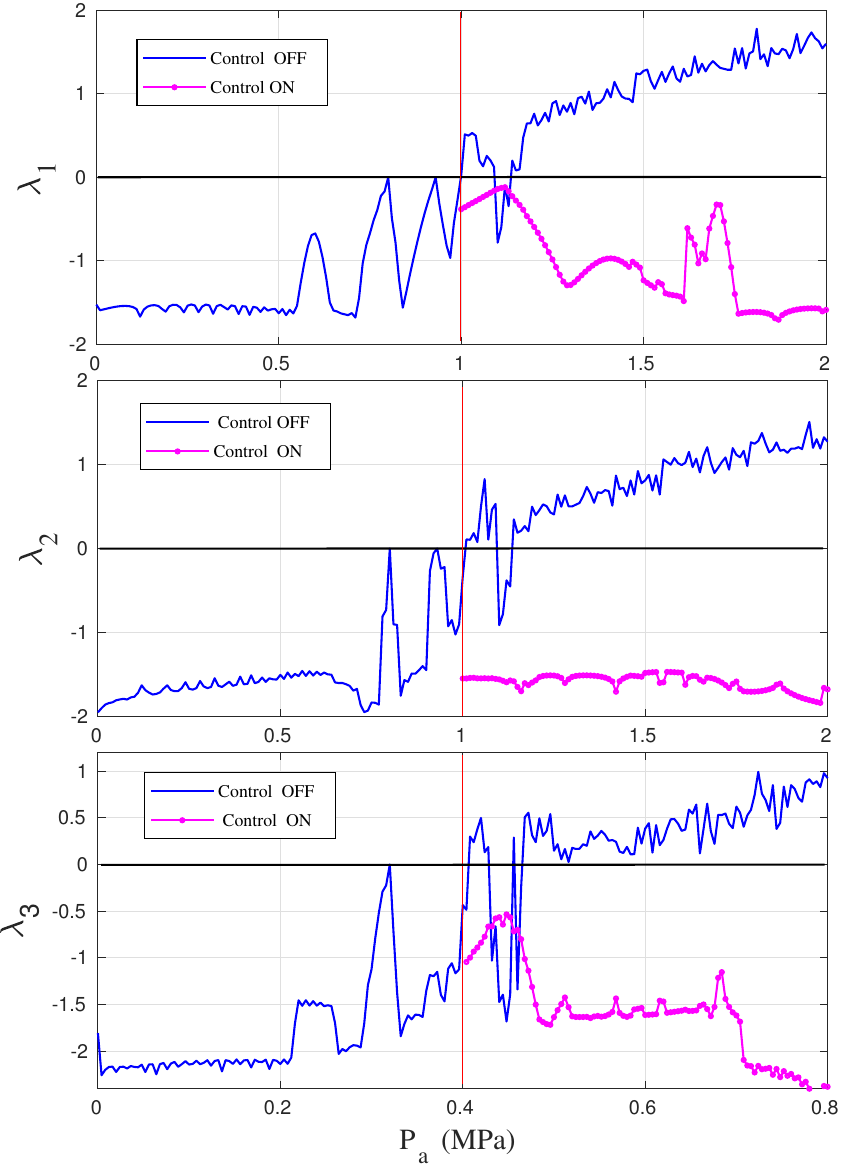}
		\caption{Lyapunov spectra of the normalized three-bubble radius driven by a $2$~MHz acoustic field with initial radii $R_{10}=4~\mu\mathrm{m}$, $R_{20}=5~\mu\mathrm{m}$, and $R_{30}=6~\mu\mathrm{m}$ as a function of pressure. The blue curve corresponds to the uncontrolled dynamics, whereas the pink curve shows the system response after applying the proposed control scheme.}\label{Fig.SB6}
	\end{center}
\end{figure}

Figure~\ref{Fig.SB2} presents the bifurcation diagrams and the corresponding maximum Lyapunov exponents for the three-microbubble interaction as functions of the driving pressure, with the acoustic frequency fixed at $3$~MHz. Transitions between different dynamical regimes are observed, including period-doubling and saddle-node bifurcations. Negative values of the maximum Lyapunov exponent indicate stable oscillatory behavior, whereas positive values signal the onset of chaotic dynamics. In comparison with Fig.~\ref{Fig.SB1}, increasing the driving frequency leads to qualitatively different transition scenarios. The results show that at higher frequencies the extent of the stable regime increases monotonically, and system trajectories are progressively attracted toward a desirable stable state. As a consequence, microbubble clusters driven at higher frequencies remain stable over a wider range of lower acoustic pressure amplitudes.

We further investigate the influence of the initial bubble radius on the collective dynamics by treating the equilibrium radius as a control parameter. Figure~\ref{Fig.SB3} illustrates the dynamical response of the microbubble cluster under an acoustic driving frequency of $1$~MHz and a pressure amplitude of $700$~kPa. As the initial radius is varied, the system undergoes successive transitions between stable and unstable regimes through period-doubling bifurcations. The emergence of chaotic windows is evident from the irregular bubble oscillations observed in the cluster dynamics. In particular, when the maximum Lyapunov exponent approaches or crosses zero [Fig.~\ref{Fig.SB3}(b)], transitions to chaotic behavior occur. Fluctuations of the Lyapunov exponent between positive and negative values reflect the intermittent alternation between stable and chaotic oscillatory states.

Although extensive numerical and experimental studies have examined bubble--bubble interactions, most existing works focus primarily on the effects of acoustic pressure and frequency, while the role of the equilibrium bubble radius has received comparatively little attention \cite{INT_21,INT_22,INT_23,INT_24,INT_25,INT_26,INT_27,INT_28,SecIV-11,SecIV-12}. Our results demonstrate that radial oscillations can become unstable or chaotic only within specific radius intervals, as shown in Fig.~\ref{Fig.SB4}. Importantly, these findings reveal that the dynamical stability of a given bubble depends sensitively on the equilibrium radii of its neighbors, highlighting the critical role of geometric heterogeneity in microbubble cluster dynamics.

\begin{figure*}
	\begin{center}
		\includegraphics[width=\textwidth]{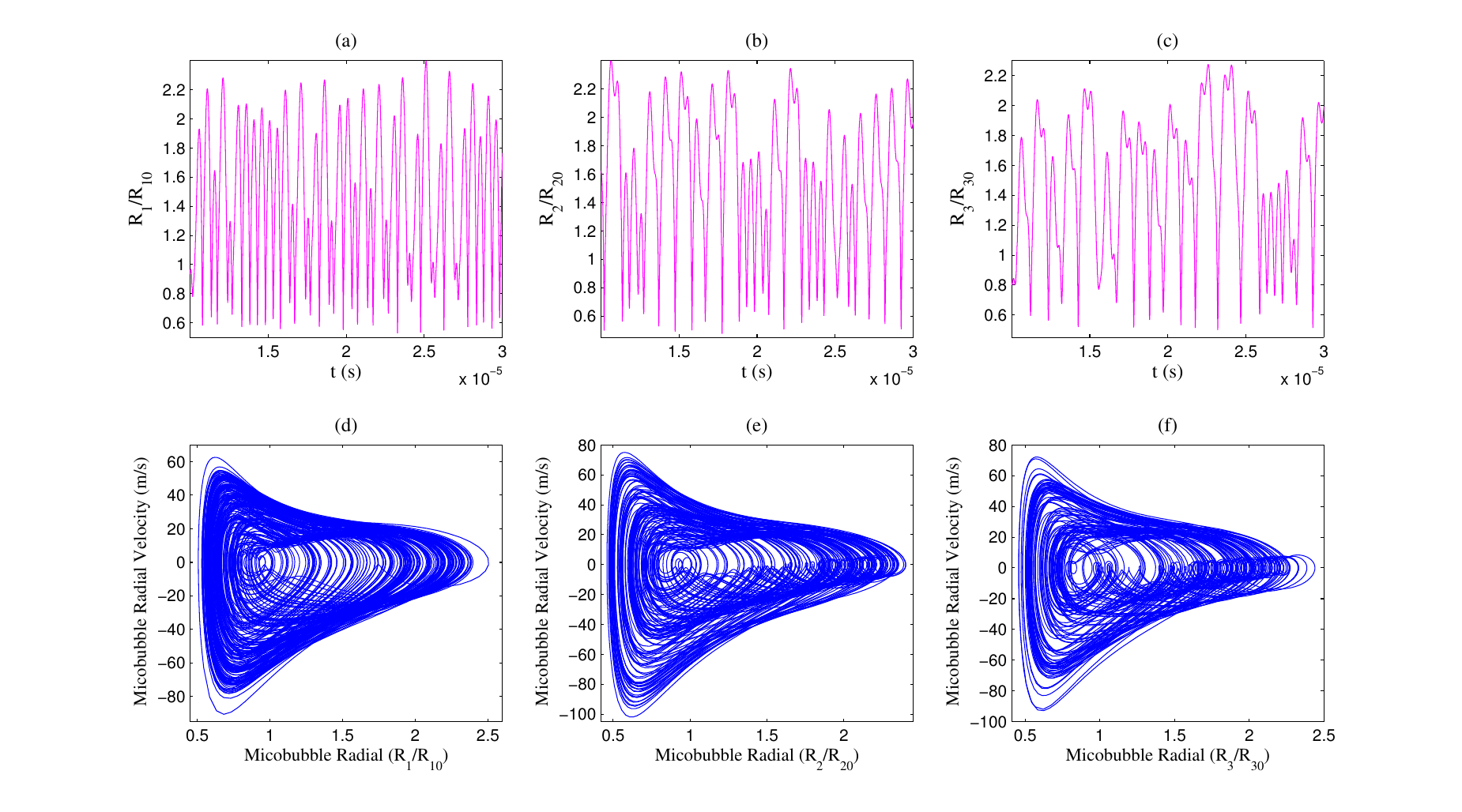}
		\caption{Time series and state-space trajectories of the normalized three-bubble radius for initial radii $R_{10}=4~\mu\mathrm{m}$, $R_{20}=5~\mu\mathrm{m}$, and $R_{30}=6~\mu\mathrm{m}$, driven at a frequency of $2$~MHz and an acoustic pressure of $1.5$~MPa, illustrating chaotic oscillations in the absence of the proposed control scheme.}\label{Fig.SB7}
	\end{center}
\end{figure*}

\begin{figure*}
	\begin{center}
		\includegraphics[width=\textwidth]{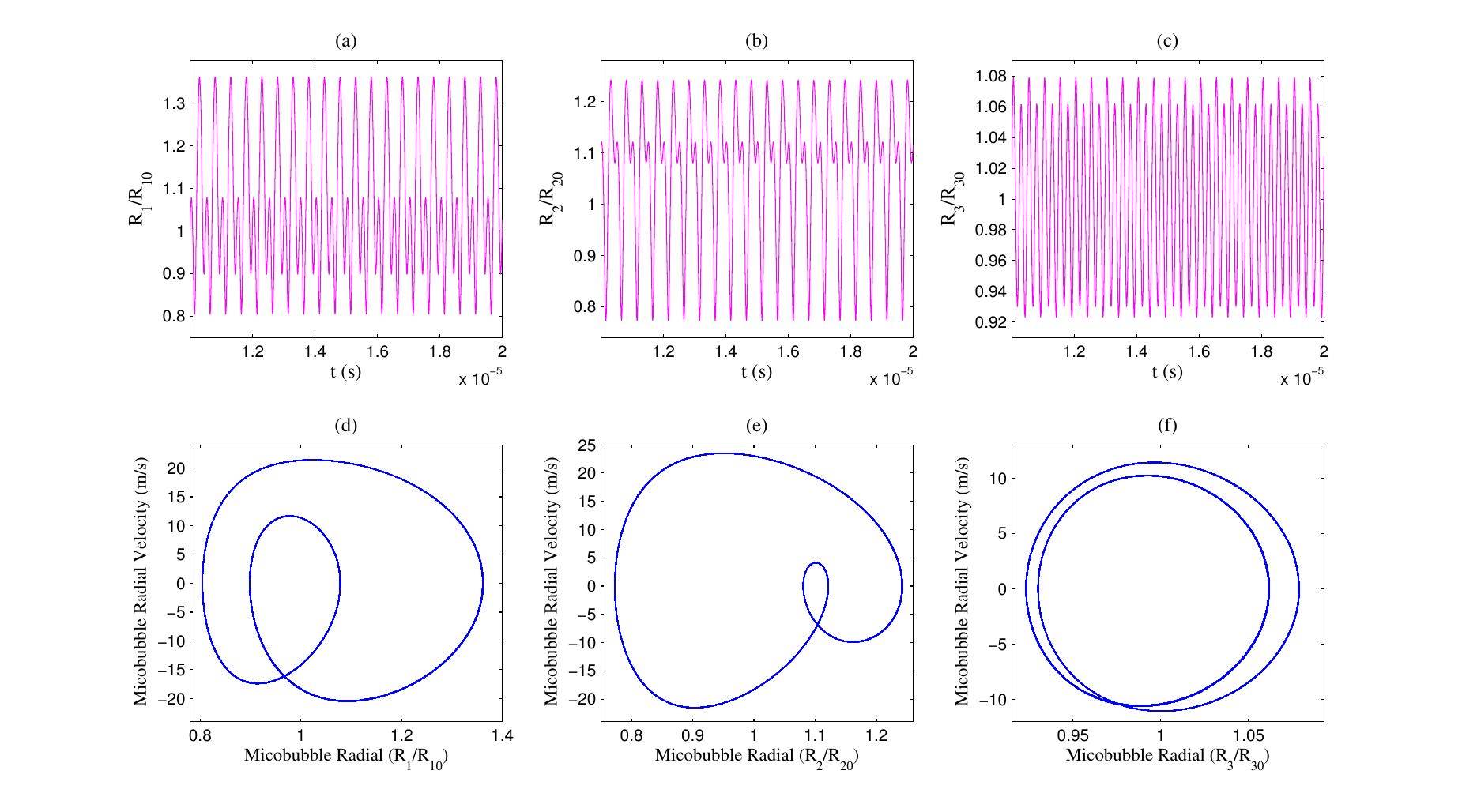}
		\caption{Time series and state-space trajectories of the normalized three-bubble radius for initial radii $R_{10}=4~\mu\mathrm{m}$, $R_{20}=5~\mu\mathrm{m}$, and $R_{30}=6~\mu\mathrm{m}$, driven at a frequency of $2$~MHz and an acoustic pressure of $1.5$~MPa, showing regular periodic oscillations after applying the proposed control technique.}\label{Fig.SB8}
	\end{center}
\end{figure*}

The chaotic regimes of the microbubble cluster are summarized in Table~\ref{Table2} for representative parameter values. These results demonstrate that the system can exhibit strongly irregular and uncontrolled dynamics depending on the chosen control parameters. Variations in acoustic pressure, driving frequency, and equilibrium bubble radius lead to qualitatively different oscillatory behaviors, including period-one motion, period-doubling transitions to period-two orbits, cascades of successive bifurcations toward chaos, high-period oscillations, and symmetry-breaking transitions. Such sensitivity to external and geometrical parameters highlights the intrinsic complexity of microbubble cluster dynamics and the difficulty of maintaining stable oscillatory behavior under practical operating conditions. To avoid undesirable dynamical responses and suppress chaotic oscillations, an effective control strategy is therefore required. In the following, we implement the adaptive intermittency-based control approach of Sec.~\ref{sec:bubblecontrol}, designed to regulate unstable bubble dynamics and restore stable collective motion.

\begin{table}[htbp]
	\begin{center}
		\caption{Domains of parameter values that lead to chaotic oscillations in the general K--H equation (for a bubble/water system at $20^\circ$C). Here $\nu$ denotes the driving frequency in MHz.}
		\label{Table2}
		\resizebox{1\columnwidth}{!}{%
		\begin{tabular}{lcccc}
			\hline\hline\noalign{\smallskip}
			Effect& Periodic domain & \multicolumn{3}{c}{Constant parameters} \\
			\noalign{\smallskip}\hline\noalign{\smallskip}
			&$P_a>0.6$     &   $\nu=1$      & $\mu=0.001$      & $R_{i0}^{\ast}=4,5,6$\\
			Pressure (MPa)&$P_a>1$       &   $\nu=2$      & $\mu=0.001$      & $R_{i0}=4,5,6$   \\
			&$P_a>1.6$     &   $\nu=3$      & $\mu=0.001$      & $R_{i0}=4,5,6$\\
			&$P_a>2.6$     &   $\nu=4$      & $\mu=0.001$      & $R_{i0}=4,5,6$\\
			\cline{2-5}\noalign{\smallskip}
			& $f<0.5$&$P_a=0.2$      & $\mu=0.001$      & $R_{i0}=4,5,6$\\
			Frequency (MHz) & $f<1$ & $P_a=0.6$    & $\mu=0.001$      & $R_{i0}=4,5,6$\\
			& $f<2$&$P_a=1$        & $\mu=0.001$      & $R_{i0}=4,5,6$\\
			& $f<3$ &$P_a=2$       & $\mu=0.001$      & $R_{i0}=4,5,6$\\
			\cline{2-5}\noalign{\smallskip}
			& $ \mu<0.002$&$P_a=1$       & $\nu=1$        &$R_{i0}=1,2,3$\\
			Viscosity (Ns/m$^2$) & $ \mu<0.005$& $P_a=1$    &  $\nu=1$       &$R_{i0}=2,3,4$\\
			& $ \mu<0.02$ &$P_a=1$   & $\nu=1$        &$R_{i0}=4,5,6$\\
			& $\mu<0.03$ &$P_a=1$       &  $\nu=1$       &$R_{i0}=6,8,10$\\
			\cline{2-5}\noalign{\smallskip}
			&  $R_{i0}<6$ &$P_a=1$       &   $\nu=2$      &  $\mu=0.001$\\
			&  $R_{i0}<10$ &$P_a=1$         &   $\nu=3$      & $\mu=0.001$\\
			Initial radius ($\mu$m)                     &  $R_{i0}<10$ &$P_a=0.4$       &   $ \nu=1$     & $\mu=0.001$\\
			&  $R_{i0}<5$ &$P_a=0.6$       &   $ \nu=1$     & $\mu=0.001$\\
			&  $R_{i0}<10$ &$P_a=0.5$       &   $\nu=2$      & $\mu=0.001$\\
			\noalign{\smallskip}
			\hline \hline \noalign{\smallskip}
			\multicolumn{5}{l}{$^{\ast}$ $i=1,2,3$.}
		\end{tabular}
	}
	\end{center}
\end{table}
	
\subsection{Chaos suppression via adaptive frequency modulation}
\label{sec:controlledbubbles}

To clearly demonstrate the efficiency of the proposed method in suppressing chaotic behavior, several representative chaotic regimes were selected as test cases. For each selected regime, the dynamical behavior of the microbubble system was analyzed both before and after the application of control. This analysis was performed by computing the bifurcation diagrams, the corresponding Lyapunov spectra, and the time-series responses. The two representative studies are defined as follows.

{\it Effect of pressure under adaptive feedback control.---}Bifurcation diagrams and Lyapunov exponents are computed for a fixed acoustic frequency of $2$~MHz and initial bubble radii $R_{10}=4~\mu$m, $R_{20}=5~\mu$m, and $R_{30}=6~\mu$m. The acoustic pressure amplitude is varied within the range of $10$~kPa to $2$~MPa. All other physical parameters are kept constant at the values listed in Table~\ref{Table1}.

{\it Effect of initial radius under adaptive feedback control.---}For fixed values of the acoustic frequency and pressure, set to $1$~MHz and $600$~kPa respectively, the influence of the initial bubble radius on the system dynamics is investigated. The initial radii of the second and third bubbles are fixed at $R_{20}=5~\mu$m and $R_{30}=6~\mu$m, while the initial radius of the first bubble, $R_{10}$, is varied within the range $1~\mu$m to $10~\mu$m. All other physical parameters are kept constant at the values listed in Table~\ref{Table1}.

In the previous subsection, our numerical simulations demonstrated that the radial oscillations of a contrast-agent microbubble can exhibit chaotic behavior. When a contrast microbubble interacts with surrounding free gas bubbles, the onset of chaos occurs at lower acoustic pressure amplitudes than in the isolated case. Moreover, the chaos threshold decreases as the number of interacting free bubbles increases, consistent with earlier reports \cite{I-31}.

The first representative example, corresponding to the pressure--bifurcation diagram of interacting microbubbles, is shown in Fig.~\ref{Fig.SB5}(a). This case considers the interaction of three microbubbles with initial radii $R_{10}=4~\mu$m, $R_{20}=5~\mu$m, and $R_{30}=6~\mu$m, subjected to a monochromatic acoustic field with frequency $f=2$~MHz. The acoustic pressure amplitude is varied within the range of $10$~kPa to $2$~MPa, which corresponds to typical conditions used in high-intensity focused ultrasound (HIFU) applications \cite{SECTIONR-2}. As the pressure amplitude increases, the stability of the microbubble dynamics is progressively reduced and chaotic oscillations emerge, in agreement with previous experimental and theoretical studies \cite{I-31,I-32}. To investigate the possibility of suppressing this chaotic behavior, the dynamical control method of Sec.~\ref{sec:bubblecontrol} is applied. The resulting controlled dynamics are presented in Fig.~\ref{Fig.SB5}(b): the extent of the chaotic region is significantly reduced after the application of the control strategy.

The maximum Lyapunov exponent provides an additional quantitative indicator of chaos in nonlinear dynamical systems. The corresponding Lyapunov spectra are presented in Fig.~\ref{Fig.SB6}. In this figure, the blue curve represents the uncontrolled system, while the pink curve corresponds to the controlled case. Positive Lyapunov exponents indicate chaotic dynamics, whereas negative values signify stable motion. The results show that the control method effectively shifts the system toward stable regimes over the entire pressure range examined.

To further illustrate the influence of control on the system dynamics, the normalized radial oscillations of the three interacting microbubbles are plotted as functions of time before and after control in Figs.~\ref{Fig.SB7} and \ref{Fig.SB8}. Figures~\ref{Fig.SB7}(a--c) display strongly irregular oscillations at the parameter set $P_a=1.5$~MPa, $R_{10}=4~\mu$m, $R_{20}=5~\mu$m, $R_{30}=6~\mu$m, and $f=2$~MHz. The corresponding phase-space projections shown in Figs.~\ref{Fig.SB7}(d--f) reveal the formation of strange attractors. After applying the control method, the system dynamics evolve toward two stable limit cycles in the state space, as illustrated in Fig.~\ref{Fig.SB8}. This behavior corresponds to a classical period-doubling bifurcation of cycles. A direct comparison between the uncontrolled and controlled cases in Figs.~\ref{Fig.SB5}--\ref{Fig.SB8} demonstrates that the proposed control strategy effectively suppresses chaotic oscillations and significantly reduces the nonlinear response of the interacting microbubble system.

\begin{figure}
	\begin{center}
		\includegraphics[width= \linewidth]{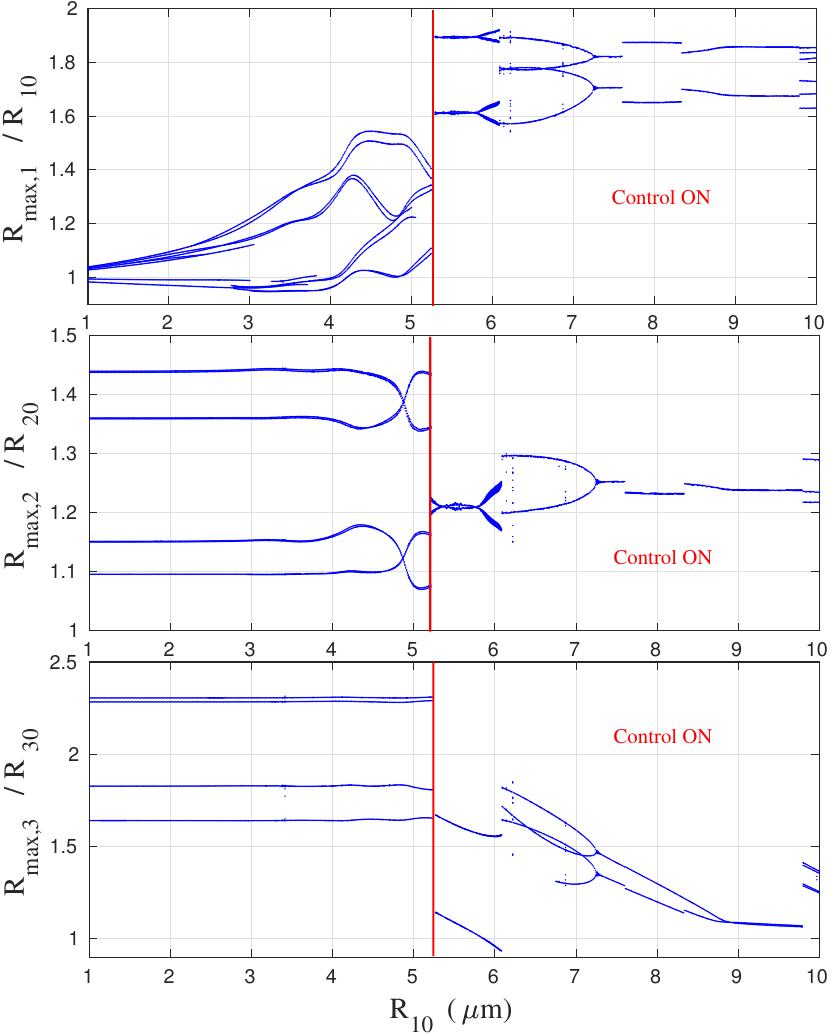}
		\caption{Bifurcation diagrams of the normalized three-bubble radius as a function of the initial radius $R_{10}$ at a driving frequency of $1$~MHz and an acoustic pressure of $600$~kPa, with fixed radii $R_{20}=5~\mu\mathrm{m}$ and $R_{30}=6~\mu\mathrm{m}$, showing the dynamical behavior of the system after applying the proposed control technique [compare Fig.~\ref{Fig.SB4}(a)].}\label{Fig.SB9}
	\end{center}
\end{figure}

By considering the initial radius $R_{10}$ as a control parameter, the influence of bubble size on the dynamics of UCA microbubbles is investigated. In clinical applications such as the treatment of uterine fibroids using high-intensity focused ultrasound, typical microbubble radii lie in the range of $1$--$10~\mu$m. Figure~\ref{Fig.SB4}(a) presents the second representative chaotic regime, corresponding to the radius--bifurcation diagram of a UCA microbubble prior to the application of the dynamical control method. In this case, the microbubble is driven by a monochromatic acoustic field with frequency $1$~MHz and pressure amplitude $600$~kPa, while the initial radius is varied as the control parameter. As the initial radius increases, the system undergoes a sequence of period-doubling bifurcations, eventually leading to chaotic oscillations. The controlled dynamics are shown in Fig.~\ref{Fig.SB9}, where a substantial reduction of chaotic behavior is observed.

Overall, the results of this section demonstrate that the proposed control strategy plays a crucial role in mitigating chaotic dynamics in interacting microbubble systems. As summarized in Table~\ref{Table3}, the application of the control method successfully suppresses chaotic oscillations over a wide range of physical parameters. The ability to suppress chaos through controlled parameter adaptation therefore offers a practical alternative to multi-frequency excitation schemes previously proposed in the literature \cite{Behnia-C,Behnia-C2,Behnia-C3,Behnia-C4,Behnia-C5,Behnia-C6}, and provides an effective route for regulating nonlinear microbubble responses in therapeutic ultrasound applications.

\begin{table}[htbp]
	\begin{center}
		\caption{Results of the control tests for parameter domains leading to chaotic oscillations in the general K--H equation (bubble--water system at $20^\circ$C). All other physical parameters were kept constant at the values listed in Table~\ref{Table1}.}\label{Table3}
				\resizebox{1\columnwidth}{!}{%
		\begin{tabular}{lccccc}
			\hline\noalign{\smallskip}
			Effect& Domain  & $P_a$ (MPa) & $f$ (MHz) & $\varepsilon$ & Result\\
			\noalign{\smallskip}\hline\noalign{\smallskip}
			&  $4<R_{0}<7$  & $1.2$          &   $2$      &  $15$&Success$^{\ast}$\\
			&  $4<R_{0}<8$ &  $1.5$         &   $2$      &  $15$&Success\\
			Initial radius ($\mu$m) &  $4<R_{0}<9$ &  $2$       &   $2$      &  $15$&Success\\
			&  $1<R_{0}<10$  &   $3$      &   $1$      &  $15$&Success\\
			&  $6<R_{0}<7$ &  $1$       &   $2$      &  $15$&Success\\
			\noalign{\smallskip}
			\hline\noalign{\smallskip}
			\multicolumn{6}{l}{\parbox{0.94\linewidth}{\footnotesize $^{\ast}$ After applying the proposed technique with the physical parameters kept constant at the values given in Tables~\ref{Table1} and \ref{Table3}, no chaotic radial oscillations occurred within $1<R_0\,(\mu\textrm{m})<10$.}}\\
		\end{tabular}
	}
	\end{center}
\end{table}

\section{Summary and conclusions}
\label{sec:conclusions}

We have introduced an intermittency-centered route to chaos suppression in which the control parameter is promoted to a dynamical variable and regulated through an auxiliary nonlinear process drawn from the same hierarchy of ergodic maps as the system itself. Using a representative one-dimensional ergodic map exhibiting intermittency without period-doubling cascades, we demonstrated that laminar--burst dynamics can be systematically controlled by dynamical coupling. The resulting feedback mechanism confines trajectories near the laminar channel, while the coupling strength $\epsilon$ provides a direct and tunable measure of stability.

The dynamics of the evolving control parameter $\alpha_m$ were analyzed explicitly: its reference map is fully chaotic and ergodic on the half line, with unstable fixed points at $\alpha^{*}=0$ and $\alpha^{*}=(2\beta+1)/\beta$ and an exact invariant density [Eq.~\eqref{eq:mualpha}], verified numerically to high precision. It is this statistically stationary evolution---converted into a bounded modulation of the map nonlinearity through the coupling function $g(\alpha)$---that suppresses intermittent bursts, rather than any convergence of $\alpha_m$ to a fixed value.

The efficiency of the control scheme was quantified using the $q$-generalized Lyapunov exponent derived from the exact SRB measure of the coupled system. The role of these Lyapunov calculations is central: they provide the analytical, initial-condition-independent order parameter of the control process, an explicit closed-form dependence of the instability on the coupling strength [Eqs.~\eqref{eq:GLE2} and \eqref{eq:GLE32}], and, through the generalized Pesin identity, the complete sensitivity to initial conditions before and after control [Eq.~\eqref{eq:xifull}]. The collapse of the positive-$\lambda_q$ regions constitutes a clear analytical signature of chaos suppression.

To establish the physical relevance of the proposed framework, we applied the same principle to interacting ultrasound-driven microbubble clusters described by the Keller--Herring formulation. By selecting the acoustically accessible driving frequency as the control parameter and allowing it to evolve dynamically according to an auxiliary nonlinear flow derived from the same conjugate-map structure, intermittent radial oscillations are progressively suppressed and replaced by stable periodic motion. Systematic bifurcation and Lyapunov analyses over wide ranges of driving pressure, frequency, and equilibrium radii---presented in full together with all model parameters and numerical procedures---confirm the robustness of the control over the experimentally relevant parameter space, including conditions typical of high-intensity focused ultrasound applications.

These results identify intermittency regulation as a unifying and physically realizable mechanism for stabilizing complex nonlinear dynamics across both abstract maps and realistic nonlinear systems. Natural directions for future work include the experimental implementation of the adaptive frequency modulation in acoustic cavitation setups, the extension of the analytical framework to higher-dimensional map hierarchies, and the application of the control principle to other systems exhibiting intermittent transitions, such as coupled oscillator networks and turbulent flows.

 \begin{acknowledgments}
 BT is supported by the Turkish Academy of Sciences (TUBA) under Grant No.
 AD-2026.
 \end{acknowledgments}

\appendix

\section{One-parameter map hierarchy: general form, invariant measures, and parameter relations}
\label{app:maps}

The maps of Sec.~\ref{sec:hierarchy} are defined in terms of Chebyshev polynomials of the first kind of degree $N$ \cite{jefm,jefm2,jefm3,jefm4,jefm5,jefm6},
\begin{equation}
	\Phi_{N}(x)=\frac{\alpha^2\left[T_N(\sqrt{x})\right]^{2}}{1+(\alpha^2-1)\left[T_N(\sqrt{x})\right]^{2}},
	\label{eq:PhiN}
\end{equation}
where $\alpha$ is the control parameter. For these maps, the invariant (SRB) measure can be obtained analytically for arbitrary $\alpha$ and integer $N$ \cite{jefm,jefm2,jefm3,jefm4,jefm5,jefm6}, and takes the form
\begin{equation}\label{eq:Mes}
	\mu_{\Phi_N}(x)=\frac{1}{\pi}\frac{\sqrt{\beta}}{\sqrt{x(1-x)}\,\left[\beta+(1-\beta)x\right]},
\end{equation}
where $\beta>0$ is a positive parameter controlling the statistical distribution of trajectories in phase space. The parameter $\beta$ is not independent, but is related to the control parameter $\alpha$ through the invariant-measure condition
\begin{equation}
	\alpha=\left\{\begin{array}{ll}
		\dfrac{\sum_{k=0}^{[(N-1)/2]}C_{2k+1}^{N}\beta^{-k}}{\sum_{k=0}^{[N/2]}C_{2k}^{N}\beta^{-k}}
		& \text{for odd } N,
		\\[14pt]
		\dfrac{\beta\sum_{k=0}^{[N/2]}C_{2k}^{N}\beta^{-k}}{\sum_{k=0}^{[(N-1)/2]}C_{2k+1}^{N}\beta^{-k}}
		& \text{for even } N,
	\end{array}\right.\label{eq:CC}
\end{equation}
where $C^{N}_{2k}$ ($C^{N}_{2k+1}$) is the binomial coefficient capturing the even (odd) indexed terms of a binomial expansion and $[\,\cdot\,]$ denotes the greatest-integer part. Considering the polynomial expansion for $N=2$, Eq.~\eqref{eq:CC} gives
\begin{equation}\label{eq:S8}
	\alpha=\frac{2\beta}{1+\beta}.
\end{equation}
Similarly, the other maps in the family satisfy, for $0<\beta<\infty$,
\begin{equation}
	\left\{\begin{array}{ll}
		\alpha=\dfrac{3\beta+1}{\beta+3} & \text{for}\quad
		\tilde\Phi_{3}(x)=\dfrac{1}{\alpha^{2}}\tan^{2}(3\arctan\sqrt{x}),\\[10pt]
		\alpha=\dfrac{4\beta(1+\beta)}{\beta^{2}+6\beta+1} & \text{for}\quad
		\tilde\Phi_{4}(x)=\dfrac{1}{\alpha^{2}}\tan^{2}(4\arctan\sqrt{x}).
	\end{array}\right.
	\label{eq:alphabeta}
\end{equation}
Substituting the map of Eq.~\eqref{eq:S7} into the relation given in Eq.~\eqref{eq:S8} yields
\begin{equation}
	\tilde{\Phi}_{2}(x)=\left(\frac{1+\beta}{\beta}\right)^{\!2}\frac{x}{(1-x)^2},
	\label{eq:tildePhi2beta}
\end{equation}
which is the form used to construct the reference map of the adaptive control scheme [Eq.~\eqref{eq:R}].

\section{Derivation of the generalized Lyapunov exponent of the uncontrolled maps}
\label{app:GLEun}

In this appendix we derive the closed-form expression for the $q$-generalized Lyapunov exponent of the maps $\Phi_N(x)$, quoted as Eq.~\eqref{eq:GLE} in the main text for $N=3$. Starting from the definition of Eq.~\eqref{eq:GLSM} and exploiting its invariance under conjugacy, the generalized Lyapunov exponent of the conjugate map $\tilde\Phi_{N}(x)$ can be written as
\begin{equation}
		\begin{split}
	\lambda_{q}&\big(\tilde{\mu},\tilde{\Phi}_{N}(x)\big)=\\ &\frac{1}{\pi}\int_{0}^{\infty}\frac{\sqrt{\beta}\,dx}{\sqrt{x}(1+\beta x)}\log_{q}\left|\frac{1}{\alpha^{2}}\frac{d}{dx}\cot^{2}\!\big(N\arctan\sqrt{x}\big)\right|.
		\end{split}
\end{equation}
Introducing the polynomials $A(x)$ and $B(x)$ defined by
\begin{equation}
	\begin{split}
		A(x)&=\sum_{k=0}^{[N/2]}C_{2k}^{N}x^{k},\\
		B(x)&=\sum_{k=0}^{[(N-1)/2]}C_{2k+1}^{N}x^{k},
	\end{split}
\end{equation}
one obtains the generalized Lyapunov exponent of the map $\Phi_{N}(x)$ in the form
\begin{widetext}
\begin{equation}
	\lambda_{q}\big(\Phi_{N}(x)\big)=\left\{
	\begin{array}{ll} \log_{q}
		\left(\dfrac{N}{\alpha^{2}}\dfrac{(\beta+1+2\sqrt{\beta})^{N-1}\prod_{k=1}^{[N/2]}(1+x_{k}^{A}\beta)}{\left(\prod_{k=1}^{[(N-1)/2]}(1+x_{k}^{B}\beta)\right)^{3}}\right)
		& \text{for even } N,
		\\[16pt]
		\log_{q}\left(\dfrac{N}{\alpha^{2}}\dfrac{(\beta+1+2\sqrt{\beta})^{N-1}\prod_{k=1}^{[(N-1)/2]}(1+x_{k}^{B}\beta)}{\left(\prod_{k=1}^{[N/2]}(1+x_{k}^{A}\beta)\right)^{3}}\right)
		& \text{for odd } N,
	\end{array}\right.
\end{equation}
where $x_k^{A}$ and $x_k^{B}$ denote the roots of $A$ and $B$. Equivalently, this result can be expressed compactly in terms of the polynomials $A(x)$ and $B(x)$ as
\begin{equation}
	\lambda_{q}\big(\Phi_{N}(x)\big)=\left\{
	\begin{array}{ll} \log_{q}
		\left(\dfrac{N}{\alpha^{2}}\dfrac{(1+\beta+2\sqrt{\beta})^{N-1}\beta^{2}A(1/\beta)}{\beta^{N-1}\left[B(1/\beta)\right]^{3}}\right)
		& \text{for even } N,
		\\[16pt]
		\log_{q}\left(\dfrac{N}{\alpha^{2}}\dfrac{(1+\beta+2\sqrt{\beta})^{N-1}B(1/\beta)}{\beta^{N-1}\left[A(1/\beta)\right]^{3}}\right)
		& \text{for odd } N.
	\end{array}\right.
\end{equation}
Finally, using the relation
\begin{equation}
	\alpha=\left\{
	\begin{array}{ll} \beta\,
		\dfrac{A(1/\beta)}{B(1/\beta)} & \text{for even } N,
		\\[10pt]
		\dfrac{B(1/\beta)}{A(1/\beta)} & \text{for odd } N,
	\end{array}\right.
\end{equation}
we obtain the closed-form result
\begin{equation}\label{eq:SMGL}
	\lambda_{q}\big(\mu,\Phi_{N}(x)\big)=\log_{q}\left(\frac{N(1+\beta+2\sqrt{\beta})^{N-1}}{\Big(\sum_{k=0}^{[N/2]}C_{2k}^{N}\beta^{k}\Big)\Big(\sum_{k=0}^{[(N-1)/2]}C_{2k+1}^{N}\beta^{k}\Big)}\right).
\end{equation}
\end{widetext}
Equation~\eqref{eq:SMGL} constitutes the general result; setting $N=3$ yields Eq.~\eqref{eq:GLE} of the main text, and the behavior near the intermittency transition follows from the limit $\beta\to0$ as described by Eq.~\eqref{eq:scaling}.

\section{Invariant measure of the coupled system: derivation}
\label{app:measure}

In this appendix we derive the factorized invariant measure of the coupled map $\Psi_N(x_m,\alpha_m)$, Eqs.~\eqref{eq:B7} and \eqref{eq:B16} of the main text, together with the explicit parameter measure $\mu(\alpha)$ [Eq.~\eqref{eq:B17} below]. For a deterministic system such as the $\Phi_{N}(x)$ map, $\Phi_{N}$-invariance implies that the corresponding invariant measure $\mu(x)$ satisfies the Frobenius--Perron (FP) integral equation \cite{Dorfmanm,32,jefm,jefm2,jefm3,jefm4,jefm5,jefm6}
\begin{equation}
	\mu(y)=\int_{0}^{1}\delta\big(y-\Phi_{N}(x)\big)\mu(x)\,dx,
\end{equation}
which is equivalent to
\begin{equation}
	\mu(y)=\sum_{x\in\Phi_{N}^{-1}(y)}\mu(x)\frac{dx}{dy}. \label{eq:B1}
\end{equation}
We define the action of the standard FP operator associated with the map $\Phi_{N}(x)$ on an arbitrary function as
\begin{equation}
	P_{\Phi_{N}}f(y)=\sum_{x\in \Phi_{N}^{-1}(y)}f(x)\frac{dx}{dy},
\end{equation}
and observe that the invariant measure $\mu(x)$ is an eigenfunction of the FP operator corresponding to the largest eigenvalue, equal to unity. Similarly, the probability measure for chaotic maps with a dynamical parameter, $\Psi_N(x_m,\alpha_m)$ defined in Eq.~\eqref{eq:PsiN}, satisfies the FP integral equation
\begin{widetext}
\begin{equation}
	\mu(x_{m+1},\alpha_{m+1})\,dx_{m+1}\,d\alpha_{m+1}=\int_{0}^{1}dx_{m}\int_{0}^{1}d\alpha_{m}\,
	\delta\big(x_{m+1}-\Phi_N(x_m,g(\alpha_m))\big)\,
	\delta\big(\alpha_{m+1}-R(\alpha_{m})\big)\,\mu(x_{m},\alpha_{m}),
\end{equation}
which is equivalent to
\begin{equation}
	\mu(x_{m+1},\alpha_{m+1})=\sum_{(x_m,\alpha_m)\,\in\, \Psi^{-1}_{N}(x_{m+1},\alpha_{m+1})}
	\left|J(x_{m},\alpha_{m})\right|\,\mu(x_{m},\alpha_{m}), \label{eq:B3}
\end{equation}
where $J(x_{m},\alpha_{m})$ is the Jacobian of the transformation,
\begin{equation}
	J(x_{m},\alpha_{m})=\frac{\partial (x_{m},\alpha_{m})}{\partial (x_{m+1},\alpha_{m+1})}
	=\frac{\partial x_{m}}{\partial x_{m+1}}\frac{\partial\alpha_{m}}{\partial\alpha_{m+1}},
\end{equation}
and the summation in Eq.~\eqref{eq:B3} is taken over all roots of Eq.~\eqref{eq:PsiN}, that is,
\begin{equation}
	x_{m,\ell,\pm}=\tan^2\!\left(\frac{1}{N}\arctan\sqrt{x_{m+1}\,g^2(\alpha_{m,\pm})}+\frac{\ell\pi}{N}\right),\quad
	\ell=0,1,2,\ldots \label{eq:B4}
\end{equation}
\begin{equation}
	\alpha_{m,\pm}=1+\frac{(1+\beta)^2}{2\alpha_{m+1}\beta^2}\pm
	\sqrt{\left(1+\frac{(1+\beta)^2}{2\alpha_{m+1}\beta^2}\right)^{\!2}+1}. \label{eq:B5}
\end{equation}
Therefore, the FP equation \eqref{eq:B3} can be written as
\begin{equation}
	\mu(x_{m+1},\alpha_{m+1})=\sum_{\alpha_{m,\pm},\;x_{m,\ell,\pm}}\left|
	\frac{\partial \alpha_{m,\pm}}{\partial \alpha_{m+1}}
	\frac{\partial x_{m,\ell,\pm}}{\partial x_{m+1}}\right|
	\mu(x_{m,\ell,\pm},\alpha_{m,\pm}).\label{eq:B6}
\end{equation}
The independence of $\partial \alpha_{m}/\partial \alpha_{m+1}$ from $x_m$ implies that the invariant measure factorizes as in Eq.~\eqref{eq:B7} of the main text, $\mu(x,\alpha)=\mu(x \mid \alpha)\,\mu(\alpha)$.

To verify that a measure of the factorized form satisfies the FP equation \eqref{eq:B1}, we take the derivatives of Eqs.~\eqref{eq:B4} and \eqref{eq:B5} with respect to $x_{m+1}$ and $\alpha_{m+1}$,
\begin{equation}
	\left\{
	\begin{array}{l}
		\dfrac{\partial x_{m,\ell,\pm}}{\partial x_{m+1}}=\dfrac{g(\alpha_{m,\pm})}{N}
		\dfrac{\sqrt{x_{m,\ell,\pm}}\,(1+x_{m,\ell,\pm})}{\sqrt{x_{m+1}}\,\big(1+g^{2}(\alpha_{m,\pm})x_{m+1}\big)},
		\\[14pt]
		\dfrac{\partial \alpha_{m,\pm}}{\partial \alpha_{m+1}}=
		\dfrac{\big(\frac{1+\beta}{\beta}\big)^2\alpha_{m,\pm}}
		{\alpha_{m+1}\sqrt{\big(4\alpha^2_{m+1}+(\frac{1+\beta}{\beta})^2\big)^2+4}},
	\end{array}\right.
\end{equation}
and substituting these into Eq.~\eqref{eq:B6} we obtain
\begin{equation}
	\begin{split}
		\tilde{\mu}(x_{m+1},\alpha_{m+1})
		&=\sum_{\ell,\pm}
		\frac{g(\alpha_{m,\pm})}{N}
		\frac{\sqrt{x_{m,\ell,\pm}}\,(1+x_{m,\ell,\pm})}{\sqrt{x_{m+1}}\,\big(1+g^{2}(\alpha_{m,\pm})x_{m+1}\big)}\,
		\frac{\big(\frac{1+\beta}{\beta}\big)^2\alpha_{m,\pm}}
		{\alpha_{m+1}\sqrt{\big(4\alpha^2_{m+1}+(\frac{1+\beta}{\beta})^2\big)^2+4}}
		\;\tilde{\mu}(x_{m,\ell,\pm}\mid \alpha_{m,\pm})\,\tilde{\mu}(\alpha_{m,\pm}).\label{eq:B9}
	\end{split}
\end{equation}
We now consider the following ansatz for the conditional measure $\tilde{\mu}(x \mid \alpha)$:
\begin{equation}
	\tilde{\mu}(x\mid\alpha)=\frac{1}{\pi}\frac{\gamma(\alpha)}
	{\sqrt{x}\,\big(1+\eta(\alpha)x\big)}.\label{eq:B10}
\end{equation}
The right-hand side of Eq.~\eqref{eq:B9} then becomes
\begin{equation}
	\frac{1}{\pi}\sum_{\pm}\left(\frac{\gamma(\alpha_{m,\pm})}
	{1+g^{2}(\alpha_{m,\pm})x_{m+1}}\sqrt{\frac{{\alpha_{m,\pm}}}{{x_{m+1}\alpha_{m+1}}}}\right)
	\sum_{\ell}\frac{1}{N}
	\frac{(1+x_{m,\ell,\pm})\,g(\alpha_{m,\pm})}{1+\eta(\alpha_{m,\pm})x_{m,\ell,\pm}}.\label{eq:B11}
\end{equation}
Using the same prescription as in Refs.~\cite{jefm,jefm2}, the final sum reduces to
\begin{equation}
	\begin{split}
		\sum_{\ell}\frac{1}{N}
		\frac{(1+x_{m,\ell,\pm})\,g(\alpha_{m,\pm})}
		{1+\eta(\alpha_{m,\pm})x_{m,\ell,\pm}}=&\;
		\frac{g(\alpha_{m,\pm})\,A_{N}\big(\eta^{-1}(\alpha_{m,\pm})\big)\big(1+g^{2}(\alpha_{m,\pm})x_{m+1}\big)}
		{B_{N}\big(\eta^{-1}(\alpha_{m,\pm})\big)}\\
		&\times\frac{1}{1+\eta(\alpha_{m,\pm})
			\Big(\frac{g(\alpha_{m,\pm})A_{N}(\eta^{-1}(\alpha_{m,\pm}))}
			{B_{N}(\eta^{-1}(\alpha_{m,\pm}))}\Big)},
	\end{split}
\end{equation}
so that Eq.~\eqref{eq:B11} reduces to
\begin{equation}
	\tilde{\mu}(x_{m+1},\alpha_{m+1})=\frac{1}{\pi}\sum_{\pm}\left(
	\frac{\gamma(\alpha_{m,\pm})}{\sqrt{x_{m+1}}}\,\frac{g(\alpha_{m,\pm})}{1+\eta(\alpha_{m,\pm})
		\Big(\frac{A_{N}(\eta^{-1}(\alpha_{m,\pm}))}{B_{N}(\eta^{-1}(\alpha_{m,\pm}))}\Big)^{2}
		x_{m+1}}\,\frac{A_{N}\big(\eta^{-1}(\alpha_{m,\pm})\big)}{B_{N}\big(\eta^{-1}(\alpha_{m,\pm})\big)}\right).
\end{equation}
On the other hand, the measure $\tilde{\mu}(x_{m+1},\alpha_{m+1})$ must itself take the form
\begin{equation}
	\tilde{\mu}(x_{m+1},\alpha_{m+1})=\frac{1}{\pi}\frac{\gamma(\alpha_{m+1})}{\big(1+\eta(\alpha_{m+1})x_{m+1}\big)\sqrt{x_{m+1}}},
\end{equation}
which is possible if $\eta(\alpha_{m,\pm})$ and $\eta(\alpha_{m+1})$ are related as
\begin{equation}
	\eta(\alpha_{m+1})=\left(\frac{g(\alpha_{m,\pm})\,A_{N}\big(\eta^{-1}(\alpha_{m,\pm})\big)}
	{B_{N}\big(\eta^{-1}(\alpha_{m,\pm})\big)}\right)\eta(\alpha_{m,\pm}),
\end{equation}
and $\gamma(\alpha_{m+1})$ is given in terms of $\gamma(\alpha_{m,\pm})$ as
\begin{equation}
	\frac{\gamma(\alpha_{m+1})}{\sqrt{\eta(\alpha_{m+1})}}=\sum_{\pm} \frac{\gamma (\alpha_{m,\pm})}
	{\sqrt{\eta(\alpha_{m,\pm})}}.
\end{equation}
\end{widetext}
Therefore, the quantity $\gamma(\alpha)/\sqrt{\eta(\alpha)}$ satisfies the FP equation associated with the map $\alpha_{m+1}=R(\alpha_m)$ and is thus proportional to its invariant measure,
\begin{equation}
	\frac{\gamma(\alpha)}{\sqrt{\eta(\alpha)}}=\tilde{\mu}(\alpha)=\frac{\sqrt{\beta}}{\pi(1+\alpha\beta)}.
\end{equation}
Finally, substituting this result into Eq.~\eqref{eq:B10}, we arrive at
\begin{equation}
	\tilde{\mu}(x,\alpha)=\frac{\sqrt{\eta(\alpha)}}{\sqrt{x}\,\big(1+\eta(\alpha)x\big)}\,\tilde{\mu}(\alpha). \label{eq:B15}
\end{equation}
Comparing Eq.~\eqref{eq:B15} with the factorized form of Eq.~\eqref{eq:B7}, and transforming back from the conjugate half-line variables to the unit interval, yields the conditional measure of Eq.~\eqref{eq:B16} together with the parameter measure
\begin{equation}
	\mu(\alpha)=\frac{1}{\pi}\frac{\sqrt{\beta}}
	{\sqrt{\alpha(1-\alpha)}\,\big[\beta+(1-\beta)\alpha\big]},
	\qquad \beta>0,\ \alpha>0,
	\label{eq:B17}
\end{equation}
which coincides with the SRB measure of the uncontrolled hierarchy [Eq.~\eqref{eq:Mes}] evaluated at $x=\alpha$.

\section{Derivation of the generalized Lyapunov exponent of the controlled system}
\label{app:GLEc}

In this appendix we evaluate the ensemble-averaged generalized Lyapunov exponent of Eq.~\eqref{eq:lambda1} and derive the closed-form result of Eqs.~\eqref{eq:GLE2}--\eqref{eq:GLE32}. Using the conjugate maps defined in Eq.~\eqref{eq:til}, the exponent can be written as
\begin{widetext}
\begin{equation}\label{eq:lambda2}
	\begin{split}
		\lambda_{q}(\mu,\Psi_{N})&=\int \mu(\alpha)\, d\alpha
		\int dx\,\mu(x\mid \alpha)\, \log_{q}\left|\frac{\partial }{\partial x} \left(\frac{1}{g^2(\alpha)}\tan^2\big(N\arctan\sqrt{x}\big)\right)\right|\\
		&=\int \mu(\alpha)\, d\alpha \int dx\,\mu(x\mid \alpha)\, \log_{q}\left|
		\frac {N\tan \big( N\arctan \sqrt {x} \big)
			\big( 1+  \tan^{2} ( N\arctan \sqrt {x})\big) }{  g ^{2}( \alpha) \sqrt {x} \,( 1+x ) }
		\right|.
	\end{split}
\end{equation}
For $N=3$, the generalized Lyapunov exponent of $\Psi_{3}(x_m,\alpha_m)$ given in Eq.~\eqref{eq:lambda2} becomes
\begin{equation}
	\begin{split}
		\lambda_q(\tilde{\mu},\tilde{\Psi}_{3})&=\int_{0}^{\infty}\tilde{\mu}(x\mid \alpha)\,
		\log_{q}\left| \frac{3}{g(\alpha_m)^{2}}\cdot\frac{1}
		{\sqrt{x}(1+x)}
		\cdot\frac{\sin 3(\arctan\sqrt{x})}
		{\cos^{3}3(\arctan\sqrt{x})}\right| dx,
	\end{split}
\end{equation}
where the $\eta(\alpha_m)/\eta(\alpha_{m+1})$ contribution integrates to zero by the invariance of the measure,
\begin{equation}
	\int_{0}^{\infty}\mu(x\mid \alpha)\,\log_{q}\big(\eta(\alpha_{m+1})\big)\,dx
	=\int_{0}^{\infty}\mu(x \mid \alpha)\,\log_{q}\Big(\eta\big(\Psi(x_{m},\alpha_{m})\big)\Big)\,dx
	=\int_{0}^{\infty}\mu(x\mid \alpha )\,\log_{q}\big(\eta(\alpha_{m})\big)\,dx.
\end{equation}
Making the change of variable $x = \frac{1}{\beta}\tan^{2}(\theta/2)$ and omitting the term $\eta(\alpha_{m})/\eta(\alpha_{m+1})$, the expression reduces to
\begin{equation}
	\begin{split}
		\lambda_q(\tilde{\mu},\tilde{\Psi}_{3})&=\frac{2}{\pi}\int_{0}^{\pi/2}d\theta\,\Big[
		\log_{q}\big((2\beta+\epsilon)+(2\beta+\epsilon)\cos\theta\big)+\log_{q}\big(\beta+\beta\cos\theta\big)\\
		&\hspace{2.2cm}+\log_{q}\Big(\tfrac{3}{4}+\cos\theta+\tfrac{1}{4}\cos2\theta\Big) +\log_{q}\big(A+B\cos\theta+C\cos2\theta\big)\Big],
	\end{split}
\end{equation}
with
\begin{equation}
	A=\frac{6\beta^2+2\epsilon\beta+3\epsilon^2}{8\beta}, \qquad
	B=\frac{2\beta^2-\epsilon^2}{\beta},  \qquad
	C=\frac{2\beta^2+2\epsilon\beta+\epsilon^2}{8\beta^2}.
\end{equation}
The above expression can be evaluated using the integral \cite{jefm,jefm2}
\begin{equation}
	\frac{1}{\pi}\int_{0}^{2\pi}d\theta\,\log_{q}\big(A+B\cos\theta+C\cos 2\theta\big)=2\log_{q}\triangle, 
\end{equation}
where
\begin{equation}
	\triangle
	=\frac{1}{\pi}\left(\frac{\sqrt{A-3C+\sqrt{(A+C)^{2}-B^{2}}}}{2}+
	\frac{\sqrt{A+B+C}-\sqrt{A-B+C}}{2}\right).
\end{equation}
\end{widetext}
Inserting the explicit expressions of $A$, $B$, and $C$ into the last integral identity and combining the four logarithmic contributions yields Eqs.~\eqref{eq:GLE2} and \eqref{eq:GLE32} of the main text.

\section{Matrix formulation of the interacting-bubble equations}
\label{app:matrix}

In this appendix we give the explicit first-order (matrix) form of the coupled Keller--Herring equations, Eq.~\eqref{eq:compactflow} of the main text. Applying Eq.~\eqref{eq:KH} to each microbubble in the cluster and performing an order-reduction procedure yields
\begin{widetext}
\begin{equation}\label{eq:kh-matrix}
	\left[
	\begin{array}{cccccc}
		1 & 0 & 0 & 0 & \cdots & 0 \\
		0 & \mathcal{A}_1(\vec{x}) & 0 & \frac{R_2^2}{d_{12}} & \cdots &  \frac{R_i^2}{d_{1i}}\\
		0 & 0 & \ddots & 0 & \cdots & \vdots \\
		0 & \frac{R_1^2}{d_{21}} & 0 & \ddots & \cdots & \frac{R_i^2}{d_{i-1,i}} \\
		\vdots & \vdots & \vdots & \vdots & 1 & 0 \\
		0 & \frac{R_1^2}{d_{i1}} & \cdots &  \frac{R_{i-1}^2}{d_{i,i-1}} & 0 & \mathcal{A}_i(\vec{x}) \\
	\end{array}
	\right]\left[
	\begin{array}{c}
		\dot{R}_1 \\
		\ddot{R}_1 \\
		\vdots \\
		\vdots \\
		\dot{R}_i \\
		\ddot{R}_i \\
	\end{array}
	\right]=\left[
	\begin{array}{c}
		\dot{R}_1 \\
		\mathcal{B}(R_1,\dot{R}_1,R_{10})-\sum_{j\neq 1}^{N_b}\frac{2\dot{R}^2_j R_j}{d_{1j}} \\
		\vdots \\
		\vdots \\
		\dot{R}_i \\
		\mathcal{B}(R_i,\dot{R}_i,R_{i0})-\sum_{j\neq i}^{N_b}\frac{2\dot{R}^2_j R_j}{d_{ij}} \\
	\end{array}
	\right],
\end{equation}
which is Eq.~\eqref{eq:compactflow} in compact notation. The functions $\mathcal{A}_i(\vec{x})$ and $\mathcal{B}(R_i,\dot{R}_i,R_{i0})$ are defined as
\begin{equation}\label{eq:AP}
	\left\{
	\begin{array}{l}
		\mathcal{A}_i(\vec{x})=\left[\Big(1-(\kappa+1)\dfrac{\dot{R}_i}{c}\Big)R_i+\dfrac{4\mu}{\rho c}+\dfrac{12\mu_{sh}\delta}{\rho c(R_i-\delta)}\right],\quad i=1,2,\ldots,N_b,    \\[14pt]
		\mathcal{B}(R_i,\dot{R}_i,R_{i0}) = \left[\dfrac{3}{2}\Big[(3\kappa+1)\dfrac{\dot{R}_i}{3c}-1\Big]\dot{R}_i^2+\dfrac{12\mu_{sh}\delta}{\rho c(R_i-\delta)R_i}\Big[\Big(\kappa+\dfrac{R_i}{R_i-\delta}\Big)\dfrac{\dot{R}_i}{c}-1\Big]\dot{R}_i
		-\dfrac{1}{\rho}\Big(\dfrac{2\sigma}{R_{i0}}+P_0\Big)\Big[1+(1-\kappa)\dfrac{\dot{R}_i}{c}\Big]\right. \\[12pt]
		\qquad +\dfrac{1}{\rho}\Big(\dfrac{2\sigma+\chi}{R_{i0}}+P_0\Big)\Big[1+(1-\kappa-3\gamma_p)\dfrac{\dot{R}_i}{c}\Big]\Big(\dfrac{R_{i0}}{R_i}\Big)^{3\gamma_p}
		+\dfrac{4\mu}{R_i\rho}\Big(\dfrac{\kappa \dot{R}_i}{c}-1\Big)\dot{R}_i+\dfrac{2\chi}{R_i\rho}\Big[(2+\kappa)\dfrac{\dot{R}_i}{c}-1\Big]\Big(\dfrac{R_{i0}}{R_i}\Big)^{2} \\[12pt]
		\qquad \left.-\dfrac{1}{\rho}\Big[1+(1-\kappa)\dfrac{\dot{R}_i}{c}\Big]P_a\sin(2\pi ft)
		-\dfrac{2\pi f R_i}{\rho c}P_a \cos(2\pi ft)\right]
		\Big/\left[\Big(1-(\kappa+1)\dfrac{\dot{R}_i}{c}\Big)R_i+\dfrac{4\mu}{\rho c}+\dfrac{12\mu_{sh}\delta}{\rho c(R_i-\delta)}\right].
	\end{array}
	\right.
\end{equation}
\end{widetext}
For the controlled system, Eq.~\eqref{eq:control-Eq}, the matrix $\tilde{J}$ is obtained from $J$ by appending a row and column with unit diagonal entry corresponding to the auxiliary frequency variable $f$, and the vector field is augmented by the row given in Eq.~\eqref{eq:Dynamical}.

\section{Methods of dynamical analysis}
\label{app:numerics}

A variety of mathematical techniques have been developed to quantify clustering and complex collective behavior, including pair-correlation functions \cite{SecV-1}, box-counting methods \cite{SecV-2}, Vorono\"{\i} tessellations \cite{SecV-3}, and Minkowski functionals \cite{SecV-4}. In the present work, the dynamical properties of interacting microbubbles are characterized primarily through the maximum Lyapunov exponent and the associated bifurcation structure. These tools are particularly suitable in situations where direct analytical approaches are not available. The maximum Lyapunov exponent, computed numerically over a wide range of control parameters, provides a clear quantitative indicator of chaotic behavior in microbubble interaction dynamics, while bifurcation analysis reveals transitions among distinct dynamical regimes, including periodic oscillations, intermittent behavior, and fully developed chaos.

\subsection{Computation of Lyapunov exponents}
\label{app:lyap}

One of the most effective approaches for characterizing the dynamical behavior of microbubble oscillations is the computation of the Lyapunov-exponent spectrum. Lyapunov exponents quantify the sensitivity of a dynamical system to initial conditions by measuring the exponential rates of divergence or convergence of nearby trajectories in phase space. They may be regarded as dynamical measures of the complexity of attractors and are defined as long-time averages of local stretching rates \cite{Dorfmanm}.

Consider two infinitesimally close trajectories in phase space with initial separation $\|\delta x_i(0)\|$ along the $i$-th direction. After a time $t$, the separation evolves to $\|\delta x_i(t)\|$. The corresponding Lyapunov exponent $\lambda_i$ is defined as
\begin{equation}
	\frac{\|\delta x_i(t)\|}{\|\delta x_i(0)\|}
	=
	2^{\lambda_i t},
	\qquad\text{i.e.}\qquad
	\lambda_i
	=
	\lim_{t\rightarrow\infty}
	\frac{1}{t}
	\log_2
	\left(
	\frac{\|\delta x_i(t)\|}{\|\delta x_i(0)\|}
	\right).
\end{equation}
Depending on the sign of the Lyapunov exponent, three distinct dynamical regimes can be identified: for $\lambda<0$, nearby trajectories converge, corresponding to stable radial oscillations; for $\lambda=0$, trajectories preserve their relative separation, indicating quasiperiodic motion on a stable attractor; and for $\lambda>0$, trajectories diverge exponentially, signifying chaotic dynamics and unstable radial oscillations. In the chaotic regime, initially neighboring trajectories separate rapidly even for infinitesimal differences in initial conditions---a phenomenon commonly referred to as sensitive dependence on initial conditions \cite{SecV-6}---and long-term prediction of the system state becomes impossible despite the deterministic governing equations.

Lyapunov exponents can be estimated from numerical time series either by direct tracking of the evolution of nearby trajectories in phase space, or by estimation based on local Jacobian matrices. The first approach is commonly known as the Wolf algorithm \cite{SecV-7} and provides an estimate of the largest Lyapunov exponent; the second allows computation of the full Lyapunov spectrum. In the present study, Lyapunov exponents are evaluated as functions of the control parameters. For each parameter value, the governing equations [Eqs.~\eqref{eq:kh-matrix} or \eqref{eq:control-Eq}] are integrated numerically, and the corresponding Lyapunov exponent is calculated. Repeating this procedure over the parameter range yields the Lyapunov spectra shown in Figs.~\ref{Fig.SB1}--\ref{Fig.SB3} and \ref{Fig.SB6}.

\subsection{Computation of bifurcation diagrams}
\label{app:bif}

Transitions from periodic motion to chaos typically occur through well-known routes such as period doubling, quasiperiodicity, and intermittency, all of which originate from local bifurcations. A bifurcation is defined as a qualitative change in the system dynamics induced by variation of a control parameter. To visualize these transitions, bifurcation diagrams are constructed for the normalized bubble radius $R/R_0$ as a function of the relevant control parameters. The analysis is performed using a Poincar\'e section, which provides a discrete representation of the continuous dynamics. In the present work, the Poincar\'e section is defined by the condition
\begin{equation}
	P \equiv \max_R\{(R,\dot{R}) : \dot{R}=0\},
\end{equation}
corresponding to the maximum bubble radius attained during each acoustic cycle. This choice has been widely employed in previous studies of cavitation bubble dynamics \cite{SecV-8}.

For the construction of bifurcation diagrams, the governing equations are integrated over $900$ acoustic cycles at the driving frequency. Only the final $300$ cycles are retained in order to eliminate transient effects. After the system reaches its steady state, up to $600$ Poincar\'e points of the normalized radius $R/R_0$ (evaluated at $\theta_0=0$) are recorded for each parameter value and plotted as discrete points in the bifurcation diagram. This procedure is repeated while gradually increasing the control parameter, thereby revealing the full sequence of dynamical transitions---from periodic oscillations to intermittency and chaotic motion---in the microbubble system. The same integration and sampling protocol is used for the controlled system, Eq.~\eqref{eq:control-Eq}, so that the diagrams before and after control (e.g., Figs.~\ref{Fig.SB5} and \ref{Fig.SB9}) are directly comparable.

\end{document}